\documentclass[a4paper,11pt]{article}
\pdfoutput=1 

\usepackage{jheppub} 

\usepackage[T1]{fontenc} 

\usepackage{subfig}
\usepackage[scientific-notation=true]{siunitx}
\usepackage{url}
\usepackage{hyperref}
\usepackage{todonotes}

\newcommand{\Secref}[1]{Section~\ref{#1}}
\newcommand{\secref}[1]{section~\ref{#1}}

\newcommand{\Tabref}[1]{Table~\ref{#1}}
\newcommand{\tabref}[1]{table~\ref{#1}}

\newcommand{\reg}{\textsuperscript{\textregistered}}
\newcommand{\tm}{\textsuperscript{\texttrademark}}

\newcommand{\nvidia}{Nvidia}
\newcommand{\poplar}{Poplar}
\newcommand{\cuda}{CUDA}
\newcommand{\gc}{Graphcore}
\newcommand{\intel}{Intel}
\newcommand{\tesla}{TESLA}
\newcommand{\xeon}{Xeon}
\newcommand{\dell}{DELL}

\newcommand{\Tf}{TensorFlow}
\newcommand{\pytorch}{PyTorch}

\newcommand{\Kmf}{K\'{a}lm\'{a}n filter}

\newcommand{\ste}{\textsuperscript{st}}
\newcommand{\nd}{\textsuperscript{nd}}

\title{\boldmath Studying the potential of Graphcore\reg{} IPUs for applications in Particle Physics}

\author[a]{Lakshan Ram Madhan Mohan,}
\author[a]{Alexander Marshall,}
\author[a,b]{Samuel Maddrell-Mander,}
\author[a]{Daniel O'Hanlon,}
\author[a]{Konstantinos Petridis,}
\author[a]{Jonas Rademacker,}
\author[b]{Victoria Rege,}
\author[b]{and Alexander Titterton}
\affiliation[a]{H H Wills Physics Laboratory, University of Bristol, UK}
\affiliation[b]{Graphcore, Bristol, UK}

\emailAdd{lakshan.madhan@bristol.ac.uk}
\emailAdd{alex.marshall@bristol.ac.uk}
\emailAdd{sam.maddrell-mander@bristol.ac.uk}
\emailAdd{daniel.ohanlon@bristol.ac.uk}
\emailAdd{konstantinos.petridis@bristol.ac.uk}
\emailAdd{jonas.rademacker@bristol.ac.uk}
\emailAdd{alexandert@graphcore.ai}
\emailAdd{victoriar@graphcore.ai}

\abstract{
This paper presents the first study of Graphcore's Intelligence Processing Unit (IPU) in the context of particle physics applications. The IPU is a new type of processor optimised for machine learning. Comparisons are made for neural-network-based event simulation, multiple-scattering correction, and flavour tagging, implemented on IPUs, GPUs and CPUs, using a variety of neural network architectures and hyperparameters. Additionally, a \Kmf\ for track reconstruction is implemented on IPUs and GPUs. The results indicate that IPUs hold considerable promise in addressing the rapidly increasing compute needs in particle physics.
}

\begin{document}
\maketitle
\flushbottom

\section{Introduction}
\label{sec:intro}
To perform high-precision measurements of rare processes, particle physics experiments require large data rates. At the Large Hadron Collider (LHC) for example, proton-proton bunch crossing rates of $40$MHz result in a typical data rate of $\mathcal{O}(1)$ TB$/$s, which must be processed in near real-time, and is expected to exceed $\mathcal{O}(10)$ TB$/$s at the high-luminosity LHC~\cite{LHCbUGIIPhysics:2018}. The future Deep Underground Neutrino Experiment is also expected to operate its data acquisition system with a throughput of $\mathcal{O}(1)$ TB$/$s~\cite{Abi:2020oxb}. Such applications currently require a large number of CPUs on site with considerable ($\mathcal{O}(1)$ PB) disk buffers. In cases where each of these events must be studied in some depth before deciding whether to save the event for offline processing, the overall signal rate is determined by the time taken to make this decision. Furthermore, these high-precision measurements require simulated data, produced `offline', that mimics the real data as closely as possible, whilst also minimising the computational burden. 

As a consequence of these constraints, many organisations within particle physics are investigating heterogeneous computing architectures as part of a strategy to cope with the vast data volumes expected in the next generation of experiments. Such architectures replace CPU-only configurations with combinations of CPUs and graphics processing units (GPUs), and sometimes additionally field-programmable gate arrays (FPGAs); see for example studies by ATLAS, COMET and LHCb~\cite{ATLASHetero:2018,Yeo:2019kna, LHCbFPGA:2020, LHCbFPGA:2020_2}. Most notably, the first level of the software trigger of the upgraded LHCb experiment will run on GPUs~\cite{Aaij:2019zbu}, and is scheduled to begin operation in 2021.

Increasingly, GPUs are also used for offline data analysis such as fitting complex theoretical distributions with many free parameters to large data samples, for example using \nvidia{}'s \cuda{} API~\cite{GooFit:2014}, or with TensorFlow based frameworks~\cite{morris:TFA, zfit:2019}. As dataset sizes in particle physics are expected to increase exponentially in the coming years, while CPU clock speeds plateau, hardware accelerators are expected become increasingly important in online and offline computing.

Over time, graphics processing units have been modified for general purpose computing workloads, and have become the dominant form of single instruction, multiple data (SIMD), accelerator hardware available to consumers. However, with the renewed interest in large-scale machine-learning (ML) algorithms, numerous machine-learning specific hardware accelerators have been developed. Recently launched by \gc{} is the \intel{}ligence Processing Unit (IPU), a new type of hardware accelerator based on a bulk synchronous parallel multiple instruction, multiple data (MIMD) architecture, and designed for machine-learning applications.

This paper represents a first investigation of the suitability and performance of IPUs in typical high energy physics ML applications, and an IPU implementation of a \Kmf. It includes benchmark tests relative to GPUs and CPUs.  The hardware used for these studies is summarised in \tabref{tab:hardware}. The code used to produce the results presented here can be found in Ref.~\cite{codeDOI}.

The paper is organised as follows: \secref{sec:IPUs}
provides an brief overview of relevant features of \gc{}'s IPUs. The subsequent sections present implementations of several particle-physics-related applications, and their performance on IPUs, GPUs and CPUs.
\Secref{sec:GANs} presents a study of generative-adversarial neural networks (GANs) for particle physics event generation and reconstruction, and in \secref{sec:DansTagging} neural network implementations for online flavour tagging. The code in these first sections is implemented in TensorFlow or pyTORCH, and can easily be executed on IPUs, GPUs and CPUs. Especially the performance differences between IPUs and GPUs are investigated in some detail for different network types and parameters.
\Secref{sec:DansKalmanFilter} explores the IPU beyond neural networks and ML, and present a Poplar-based implementation of a \Kmf, one of the most ubiquitous track reconstruction tools in particle physics. Finally \secref{sec:conclusion}, concludes this paper.

This research was carried out by particle physicists at the University of Bristol, in close collaboration with \gc{} who provided cloud access to their IPU server as well as software support. One of the university team's PhD students became \gc{} employee during this collaboration.

\section{Graphcore's IPU}
\label{sec:IPUs}
\noindent The IPU is a new type of processor designed specifically for ML applications.  Its architecture is fundamentally different to that of either CPU or GPU. A detailed review of the architecture and performance of the first generation IPUs used in this paper can be found in Ref~\cite{jia2019dissecting}.

The IPU processor is optimised to perform highly-parallelised fine-grained operations. In contrast to the Single Instruction, Multiple Data (SIMD) architecture of GPUs, which requires contiguous vectorised data for efficient operation, the IPU is highly efficient on applications that require irregular and sparse data access and can run individual processing threads on small data blocks while exploiting its Multiple Instructions, Multiple Data (MIMD) architecture.

\begin{figure}[h!]
\centering
\includegraphics[width=1.0\textwidth,keepaspectratio]{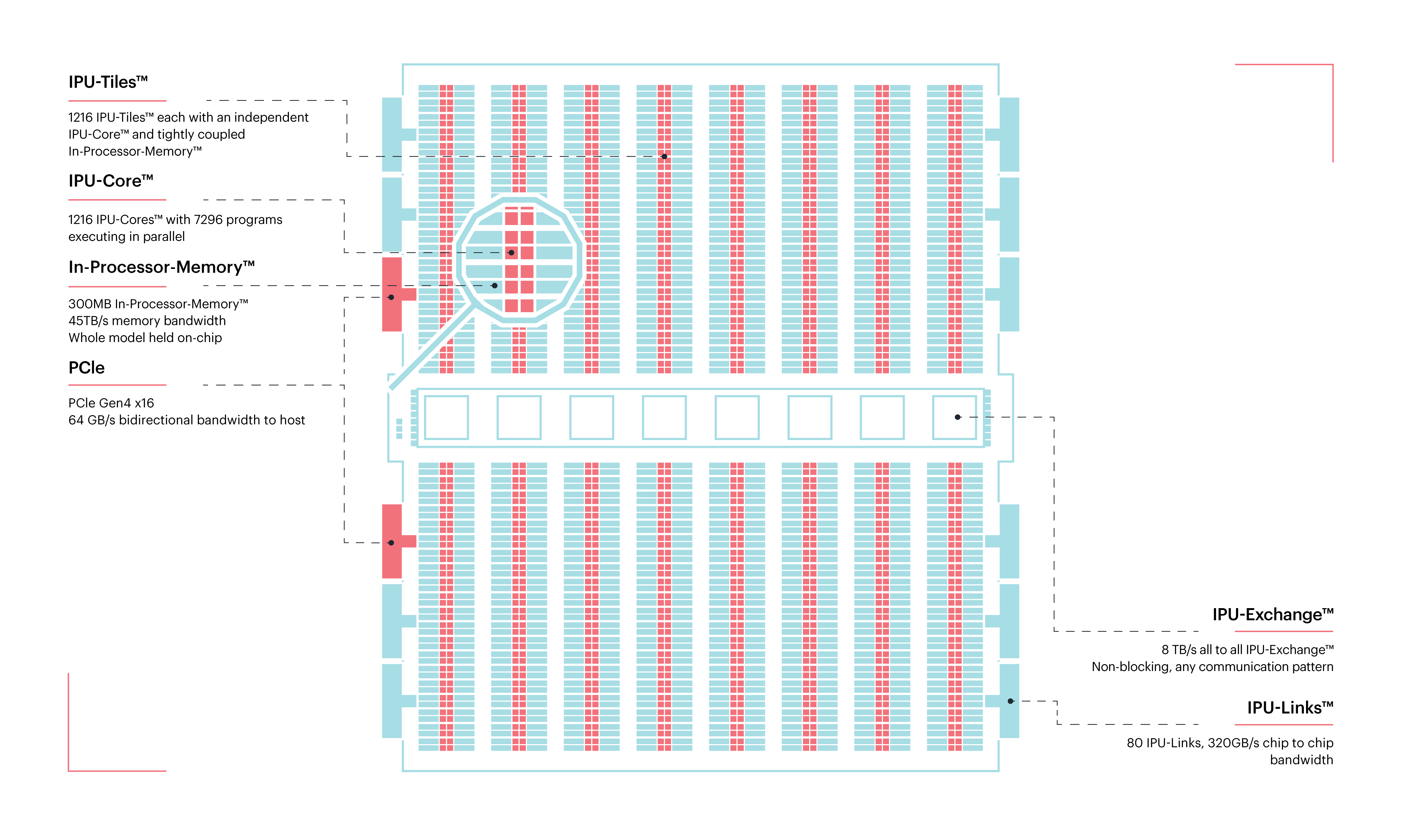}
\caption{The \gc{} Colossus\tm~MK1~GC2 IPU~\cite{gc_weblink}.
\label{ipu}}
\end{figure}

This study makes use of \gc{}'s first generation Colossus\tm~MK1~GC2 IPU (see Figure~\ref{ipu}). This IPU comprises 1,216 processing elements, called tiles, each of which consists of a computing core with 256KiB of local memory. In total 7,296 threads can be executed in parallel in a single IPU. The tiles are linked through an on-chip interconnect, the IPU~exchange\tm{}, allowing for a low-latency and high-bandwidth communication up to 7.7~Tb/s. Each IPU card consists of two such IPUs. The IPUs are connected to each other via 80 IPU~links\tm{} reaching a total chip-to-chip bandwidth of 2.5~Tb/s, and are connected to the host via 16 PCIe~Gen4 links (8 per IPU).

The IPUs used here are integrated in a \dell{} DSS8440 IPU server containing 8 dual IPU cards. This server includes two \xeon{} Platinum 8168 CPUs with 24$\times$32GB 2.4GHz DDR4 DIMM Modules. \gc{} also provides drivers along with its \poplar{} Software Development Kit (SDK). Updates to both the drivers and SDK can result in improvements to the IPU perfmance. This paper relies on SDK version v1.2.0.

During the preparation of this paper, \gc{} released its second generation IPU, the Colossus\tm~MK2~C200 with 20\% more tiles and triple the local memory per tile~\cite{gc_weblink}.

In this paper the performance of a single first generation IPU is tested against a $\textrm{\nvidia{}~\tesla{}~P100}$ GPU and two types of CPUs, depending on the particular form of the test. The power consumption of the single IPU is approximately half that of the GPU.
Key technical specifications of the IPUs, GPUs and CPUs used are given in Table~\ref{tab:hardware}.

\begin{table}
    \caption{Key specifications of the processors used in this paper as provided on manufacturer websites~\cite{intel_weblink_1, intel_weblink_2, nvidia_weblink,gc_weblink}, and in~\cite{gc_personal, jia2019dissecting}. Many features are not represented in this table; key differences in performance arise from the very different memory architectures and technologies. Performance in terms of floating point operations per second (FLOPS) is given for 32 bit single-precision operations. Thermal design power (TDP) is given for each processor, where for the IPU this is half of the total board TDP.
    \label{tab:hardware}
    }
    \centering

    \begin{tabular}{ l| *{5}{r|}}
              & Name & Cores & Memory & Clock Speed & TDP
              \\\hline
        CPU 1 &  \intel{} \xeon{} Platinum 8168 & 24 & 732 GiB & 2.7 -- 3.7 GHz            & 205~W \\
        CPU 2 & \intel{} \xeon{}  E5-2680 v4&    14 & 128 GiB & 2.4 -- 3.3 GHz & 120~W  \\
                \multicolumn{6}{c}{}\\
              & Name & Cores & Memory & 32 bit FLOPS& TDP \\
              \hline
        GPU   &  \nvidia{} \tesla{} P100 &    3584 & 16000~MiB & 9.3 TFLOPS& 250~W \\
        IPU   &  \gc{} Colossus\tm~GC2 &     1216 & 286~MiB & 31.1 TFLOPS & *120~W
    \end{tabular}

\end{table}

IPUs out-perform GPUs in many machine-learning applications such as computer vision, natural language processing and probabilistic modelling \cite{graphcore_benchmarks_2020, matthew_graphcore_2020, masters_graphcore_2020}. Machine learning has been used in particle physics for decades, initially referred to as `multivariate analysis' and typically carried out with tools developed by and for particle physicists, such as the widely-used TMVA package~\cite{TMVA:2009}. Increasingly, though, industry-standard tools and environments are being used, such as \cuda{} \cite{CUDA} TensorFlow \cite{tensorflow2015-whitepaper} and PyTorch \cite{paszke2017automatic}.
While ML algorithms are most frequently applied in the final stage of event selection, they are also used for particle identification~\cite{LHCbPerformance:2014}, flavour tagging~\cite{Aaij:2016psi} and triggering~\cite{LHCbTrigger:2012, BonsaiBDTTrigger:2012}. Neural networks have been studied for use in track reconstruction~\cite{Rinnert:2019vxw}, motivated by their high performance on hardware accelerators like GPUs and FPGAs.

The increased use of GPUs in particle physics offline data analysis coincided with the advent of increasingly user-friendly programming environments (such as \cuda{} and TensorFlow), that allow programmers without special training to fairly quickly gain access to GPU programming. Such environments exist for IPUs already, including TensorFlow, PyTorch, and \gc{}'s C++ based API, Poplar. Ease of programming is a substantial advantage over FPGAs, and is, apart from performance, a key reason that motivates our study of potential use of IPUs in particle physics.

In the same way as GPUs outperform CPUs not only in the graphics applications they were originally designed for, but also other applications such as ML, it is reasonable to expect IPUs to excel in applications beyond ML; particularly promising are those that benefit from the IPU's flexible MIMD architecture, that contrasts with the GPUs SIMD design.

\section{Event Generation and tracking corrections using GANs}
\label{sec:GANs}

    Generative Adversarial Networks (GANs) are a class of flexible neural network architectures characterised by a two-player adversarial training environment where the response of a classification discriminator network informs the updates to a generator network~\cite{goodfellow2014generative}. The discriminator is trained to distinguish between generated samples and samples from a training set. The generator network transforms a vector of random noise into a fabricated sample. GANs are trained with an iterative approach, this allows the generator and discriminator networks to improve together in parallel. The goal of GAN training is to create a generator that is able to emulate the characteristics of a training data set with high fidelity.

    In the ML community GANs have been shown to work well across a spectrum of tasks. The most common task is the generation of data in the form of images~\cite{karras2017progressive, yu2018generative, zhang2019self, zhang2017stackgan}. Increased functionality in the GAN comes with the introduction of conditional inputs into the generator, where the conditional arguments represent characteristics of the generated sample. The conditional input could be an input image to which a style transfer can be applied~\cite{Isola2016ImagetoImageTW}, or the resolution upscaled to reconstruct sub-pixel information~\cite{Ledig2017PhotoRealisticSI}. The flexibility of neural networks enable the creation of a wide range of architectures. These recent developments in the ML community, catalysed by hardware improvements, have improved generative neural networks to the point that they can feature as viable tools within particle physics computation. GANs are capable of modelling high dimensional distributions or transformations and are able to generate samples with high fidelity to training information. Conditional architectures can be designed to enable the networks to understand physical processes.

    Applications of GANs within particle physics are constantly appearing. GANs have been applied in both event generation~\cite{de2017learning, ahdida2019fast, di2019dijetgan, Butter:2019eyo, Martinez:2019jlu, Carrazza:2019cnt, Butter:2019cae} and detector modelling~\cite{paganini2018calogan, paganini2018accelerating, Maevskiy:2019vwj, Erdmann:2018jxd, Buhmann:2020pmy, Bellagente:2019uyp, Ghosh:2020kkt, Carminati:2020kym, Belayneh:2019vyx}. In this section the inference and training speeds of some of these particle physics based GANs are assessed on the IPU hardware and compared to results on the GPU and CPU described in Table~\ref{tab:hardware}.

\subsection{Event Generation}\label{Alex_event_generation}

    Accurate event generation is a crucial component of modern particle physics experiments. Large samples of simulated particle physics processes, including the detector response, are required in order to optimise the design of the detectors, develop reconstruction algorithms, understand the efficiency sub-systems and model the impacts of various physics based selection criteria. Experiments at the LHC simulate billions of events every year, each event taking up to $\mathcal{O}(\textrm{min})$ to simulate ~\cite{de2017learning}. This results in simulation campaigns consuming up to $70\%$ of experiment computing resources ~\cite{karavakis2014common,paganini2018calogan}.

    Newly proposed experiments will continue to demand a rapid increase in the number simulated events~\cite{apollinarihigh,anelli2015facility}. The ongoing optimisation and parallelisation of traditional event generation software will at best result in an order of magnitude reduction of resources~\cite{Canal:2016dki,Amadio:2020ink}. This reduction is not sufficient to meet ever increasing simulation demand. Estimates forecast a 4-fold shortfall of computing power within the next 10 years without significant new investment~\cite{albrecht2019roadmap,musella2018fast}. This has catalysed efforts to develop faster simulation and event generation technologies of which GANs are currently a front runner. GANs or other generative network architectures are likely to become an integral part of a future fast simulation tool kit.

    GANs are, of course, unable to completely replace traditional simulation methods as they rely on training data produced with the slower full physics simulation, this fact makes the optimisation of traditional methods no less valuable.
    GANs learn by example and are largely limited to modelling the exact process that they were trained on.
    In comparing a GAN to the full simulation care needs to be taken to assign a systematic uncertainty related to the residual mismodelling. The GAN event generation is particularly helpful when the systematic uncertainty due to its mismodelling is smaller than other errors associated with other parts of the analysis procedure~\cite{ahdida2019fast}. A limitation of the GAN-based event-generation stems from the fact that the range of the feature space that the GAN can accurately model is defined by that of the full-simulation training sample. However, GANs are able to accurately interpolate between points in the feature space of the training sample, acting as a powerful data augmentation tool.

    Using GPUs to generate events using a GAN-based approach offers large increases in event-generation rate over traditional simulation approaches~\cite{ahdida2019fast, de2017learning, Erdmann:2018jxd}. However further increases in the rate would be valuable. This section investigates if IPUs can provide any additional increase in the inference speed of a GAN for event generation.

    Examples of GAN architectures are taken from the literature and event-generation rates are compared across a range of batch sizes and different hardware options. Currently, convolutional networks are the most commonly used in the particle physics community.
    Two such networks are investigated here, the small convolutional DijetGAN from~Ref.~\cite{di2019dijetgan} and the larger locally connected LAGAN from~Ref.~\cite{de2017learning}.
    Additionally, two fully connected networks are investigated. These are the prompt and non-prompt muon kinematic generators developed for the SHiP experiment in~Ref.~\cite{ahdida2019fast}.
    Both fully connected networks are of similar architecture, however the prompt network is significantly smaller. As the network weights are not publicly available for all the network architectures under study, random values are assigned to the network weights without affecting the speed of the event generation.

    \begin{figure}[ht]
    \centering
    \includegraphics[width=0.99\textwidth,keepaspectratio]{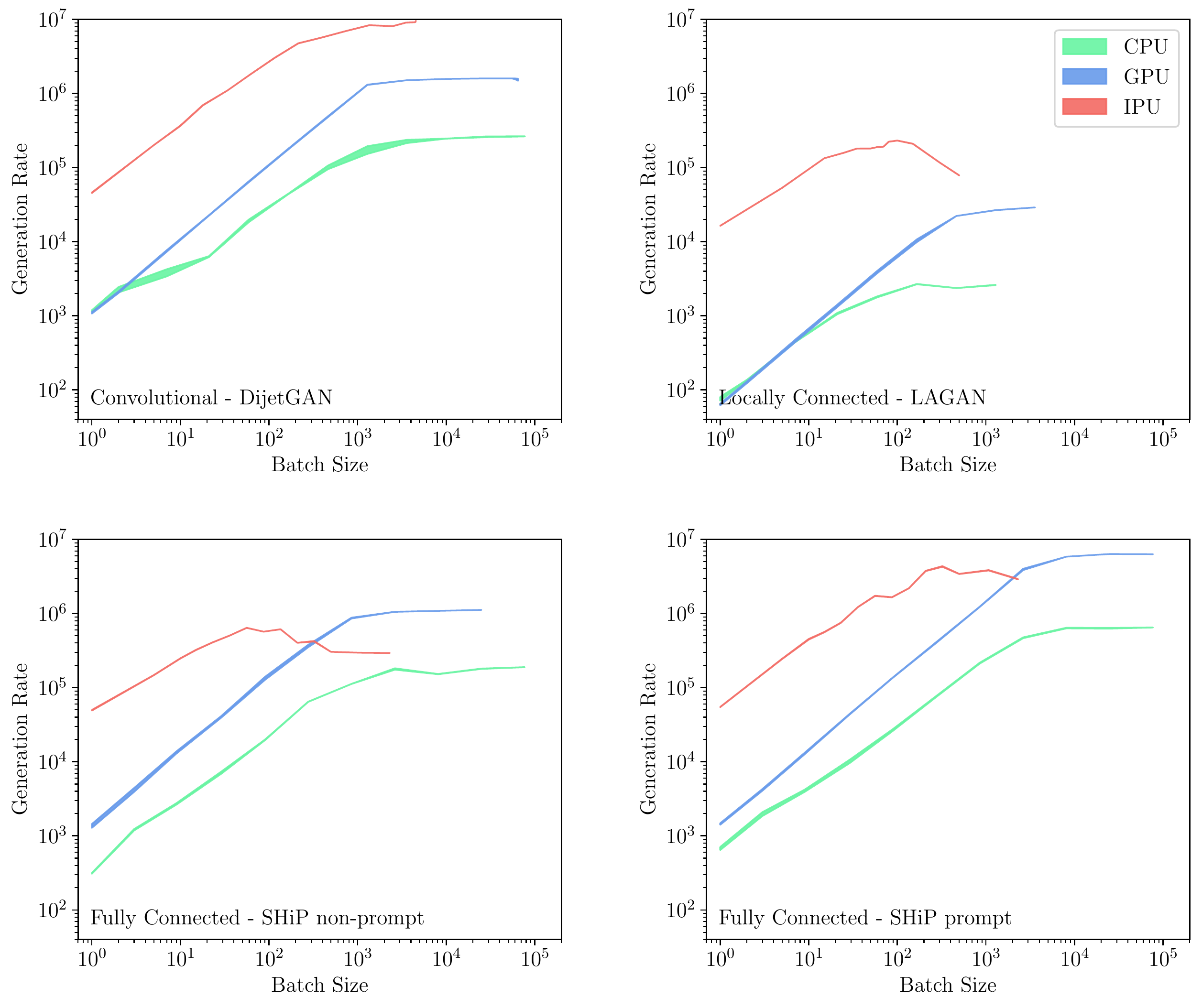}
    \caption{
    Benchmarking results of the event-generation rate  as a function of the batch size of the network. Results are presented for IPU, GPU and CPU hardware options outlined in Table~\ref{tab:hardware}.%
    \label{Alex_generation_results}}%
    \end{figure}

    Figure~\ref{Alex_generation_results} presents the event-generation rate for CPU, GPU and IPU as a function of the batch size for of the each networks studied. The relationship between rate and batch size is shown to be consistent across network and hardware configurations, with larger batch sizes giving larger generation rates. However, there is a limit to the maximum batch size accessible by each hardware option due to memory constraints. This limitation results in a plateau of the event-generation rate.

    For each network architecture and hardware option, the batch size that gives the largest event-generation rate is chosen. The CPU and GPU results are obtained with \Tf{}~\texttt{2.1.0} and the IPU results are obtained using \Tf{}~\texttt{1.15.0} as \gc{}'s SDK version 1.2.0 offered a more comprehensive support for \Tf{}~\texttt{1.x}.

    \begin{table}[]
    \caption{Benchmarking results calculated using optimal batch size for each hardware option.}
    \centering
    \resizebox{0.85\textwidth}{!}{%
    \begin{tabular}{r|l|l|l}
    Network Name    & Number of Parameters & IPU/CPU rate & IPU/GPU rate \\
    \hline
    DijetGAN        & \num{3e4} & 36.3             & 6.0              \\
    LAGAN           & \num{4e6} & 86.5             & 8.0              \\
    SHiP non-prompt & \num{5e6} & 3.4              & 0.6              \\
    SHiP prompt     & \num{6e5} & 6.7              & 0.7
    \end{tabular}%
    }
    \label{tab:Alex_table}
    \end{table}

    Across all networks tested the IPU is faster than the GPU at generating events using small batch sizes. For the fully connected networks, both of which have 2 hidden layers, the GPU becomes more efficient at higher batch sizes which are not accessible by the IPU that was used due to memory constraints. As the batch size approaches the limit for a single IPU, the performance appears to degrade. This is most likely due to overheads in the computation associated with organising large tensors in memory. At the most efficient point, the fully connected networks were $1.4$ and $1.7$ times faster using the GPU for the smaller
    and larger networks respectively.

    In contrast, the IPU outperforms the GPU for both of the convolutional networks tested. For optimal batch sizes, the IPU presents an increase in event-generation rate compared to the GPU by a factor of $6.0$ and $8.0$ for the small and large networks respectively.

\subsection{Track corrections}\label{sam_srGAN}

The use of GANs extends beyond event generation and can be employed in data processing. Charged particles traversing a medium are deflected through multiple small-angle scatters due to the Coulomb force acting between the charged particle and the nucleus of the material. The resulting trajectory of the particle is therefore modified by this scattering and traditional tracking methods rely on techniques such as the K\'{a}lm\'{a}n Filter, discussed in Sec.~\ref{sec:DansKalmanFilter}, to account for this effect. Such methods can be computationally expensive. Therefore, employing a fast pre-processing stage prior to the track-fit that corrects for the effects of multiple scattering could be desirable.

Previous work on GANs has shown that in addition to conditional class information, a generator can be conditioned with an input state to be manipulated. This is typically an input image to which a style transfer can be applied~\cite{Isola2016ImagetoImageTW}, or the resolution upscaled to recover sub-pixel information~\cite{Ledig2017PhotoRealisticSI}.
This family of transformations is of particular interest in particle physics and other scientific domains, as it shows that by using a GAN high fidelity information can be correctly recovered. In the context of particle physics, this could mean correcting for the resolution of the detector, accounting for detector misalignment or upscaling the reconstructed hit information of charged particles to correct for effects such as multiple scattering prior to a track fitting algorithm.

In order to provide a simple concrete example, the algorithm presented in this paper aims to correct for the effect of multiple scattering from the trajectory of a charged particle in two dimensions. A simplified simulation is developed to model the multiple scattering of a charged particle traversing a series of active detection material made of silicon. The multiple scattering of the charged particle with each layer of silicon is modelled according to Ref~\cite{PDG2018}, where the particle's path is deflected according to a Gaussian distribution whose width depends on the original particle's momentum and velocity as well as the thickness of the scattering medium. The same initial conditions are used to generate a second, `true',  charged particle that does not undergo scattering. The GAN is trained to perform a style transform from the scattered track to true track.

The generator model used for this study is based closely on the \texttt{pix2pix} algorithm~\cite{Isola2016ImagetoImageTW} as it has been shown to generalise over different applications without major changes to the network architecture.
The generator model consists of a U-Net encoder-decoder structure  ~\cite{Ronneberger2015UNetCN} with  ``skip" layers between each of the layers. The skip connections allow to scale specific information to directly pass across the generator and bypass the bottle neck.
The key difference to GANs used for image generation is an additional super resolution layer to upscale the output. The variation of this model used to model charged tracks is referred to as qSRGAN.

\begin{figure}
    \centering

    \includegraphics[width=0.99\textwidth]{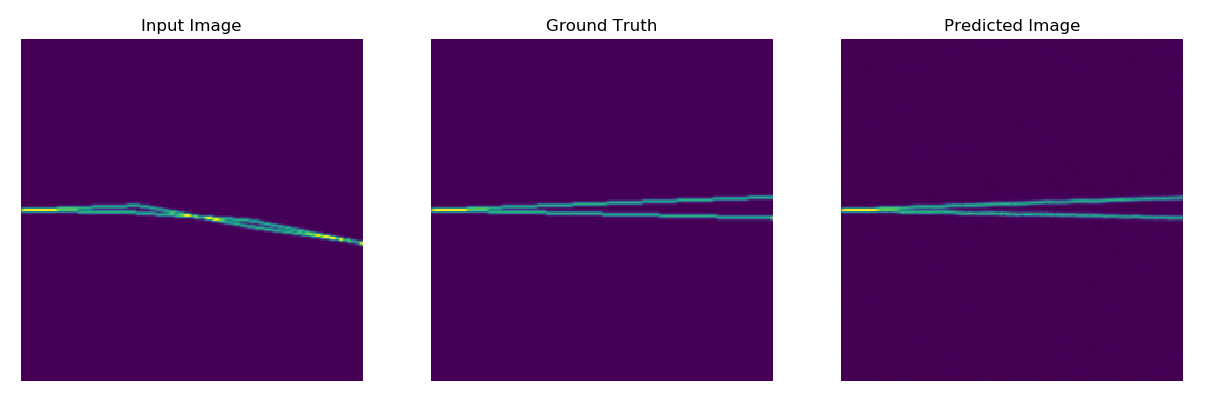}
    \caption{An example of correcting for the track multiple scattering using the qSRGAN. The left image is the input to the Generator, the middle image is the true image with no scattering, and the right image is the generated output.}
    \label{fig:tracks_recon}
\end{figure}

An example of how this algorithm performs on a pair of tracks is shown in Fig.~\ref{fig:tracks_recon}.

\begin{figure}
    \centering
    \includegraphics[width=0.6\textwidth]{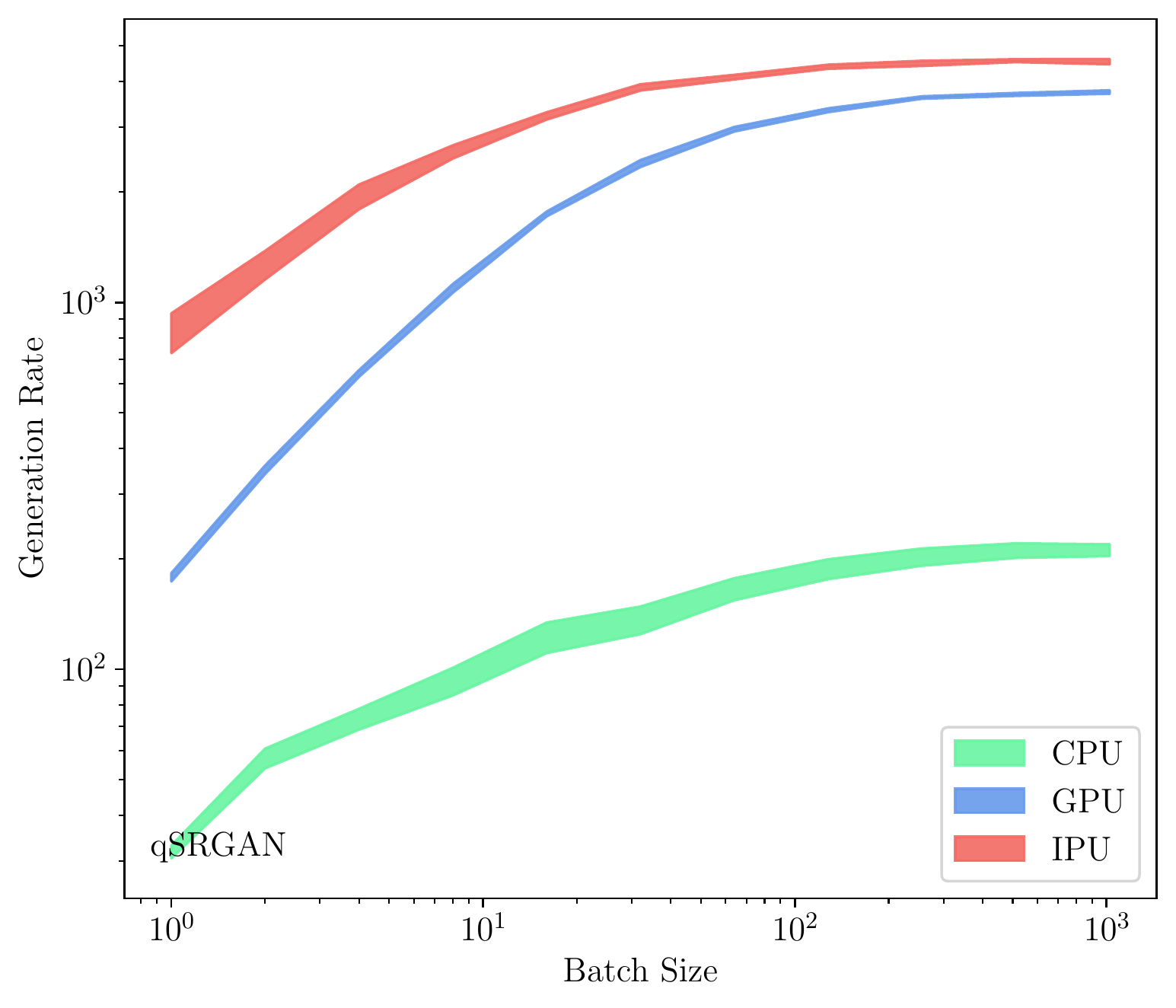}
    \caption{Benchmarking the qSRGAN algorithm on CPU, GPU, and IPU processors. The inference throughput for each processor is shown as a function of batch size (left), and the relative ratios shown for the design batch size of one image (right).}
    \label{fig:IPU_gan_sam}
\end{figure}

In contrast to event generation methods described in Sec.~\ref{Alex_event_generation} where the maximal throughput is obtained using larger batches, track corrections would typically be done on and event-by-event basis. This allows the performance of the IPU at low batch size to be utilised efficiently.
The performance of the qSRGAN algorithm for inference is tested on the CPU, the GPU and the IPU given in Table~\ref{tab:hardware}. Two key results are presented. Firstly the throughput of the algorithm as a function of batch size, and secondly the ratio of the rates of the CPU and GPU to the IPU for a batch size of one image. The results are shown in Fig.~\ref{fig:IPU_gan_sam} where the rate of the image generation using an IPU is larger by a factor of 22 relative to a CPU, and 4.5 relative to the GPU.
The increased generation rate of the IPU compared to the GPU would allow either a higher total throughput to better cope with higher event rates, or a significantly more complex model for the same total compute budget.

\subsection{Training Models}\label{Alex_training_section}

The results of Secs.~\ref{Alex_event_generation}~and~\ref{sam_srGAN} show that IPUs outperform GPUs for networks with a small batch size. Trained GANs used for event generation are implemented using the optimal batch size, which generally corresponds to the largest batch size accessible to the hardware. However, a small batch size whilst training the GAN results in more updates of the network configuration for the same computing power. Each update also contains a stochastic component originating from the random selection of training samples. This stochastic effect can help to move network configurations out of local minima. Larger batch sizes have advantages too, more efficient computation per training sample and a more accurate assessment of the gradient at each step. So called mini-batch gradient descent aims to operate with a batch size that balances this stochastic effect with the accuracy of gradient updates computed with large batch sizes. Appropriate choice of the batch size during training of the network can provide a faster overall convergence to an optimal configuration. Commonly the batch size chosen for training a GAN is $\mathcal{O}(50)$.

\begin{figure}[h!]
\centering
\includegraphics[width=0.9\textwidth,keepaspectratio]{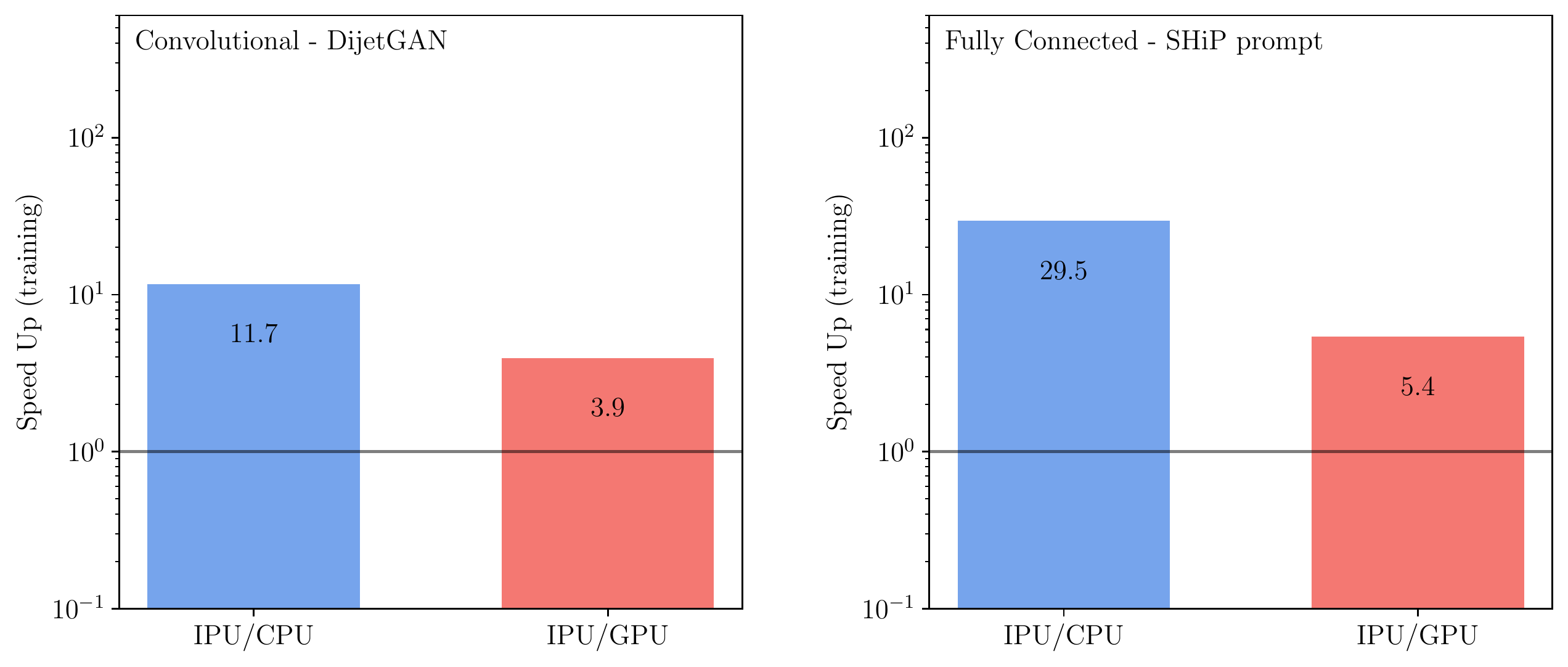}
\caption{Comparison of the time to train the IPU relative to the CPU or GPU of Table~\ref{tab:hardware}.}%
\label{Alex_training_results}%
\end{figure}

This section investigates the performance of the IPU for training the GANs described in Sec.~\ref{Alex_event_generation}.
The smaller models of the dijetGAN and SHiP prompt GAN, are trained on a single IPU. The training time is defined as the time taken to run over 1000 batches using the batch sizes reported in their respective publications. The batch sizes are 50 for the SHiP prompt GAN and 128 for the dijetGAN. The IPU training times are then compared to the same test completed on the GPU and CPU from Table~\ref{tab:hardware}. The results are presented in Fig.~\ref{Alex_training_results}. Both networks train significantly faster on the IPU as expected from the inference performance discussed in Secs.~\ref{Alex_event_generation}~and~\ref{sam_srGAN}, where for lower batch sizes the IPU consistently outperforms the GPU.

\section{Determining the flavour of $B$ mesons}
\label{sec:DansTagging}
Neural networks are commonly used to combine lower-level detector-specific information to determine the identity or quark composition of a particle. Given the large number of particles produced in each collision event, inference speed is an important consideration, regardless of whether these are evaluated `online' as part of the reconstruction and trigger framework, or `offline' after the initial rate reduction from the trigger.

For some applications, such as the determination of the flavour of the $B^0_{(s)}$ meson at production time, significantly increased classification accuracy is achieved by applying a network over all particles in the event, rather than selecting particles thought to be of particular interest ahead of time. In this way, correlations between the features of different particle tracks can also inform the resulting flavour determination. Two canonical neural network components that enable this multidimensional data to be taken into account are convolutional and recurrent neural networks. In general, recurrent networks are able to better exploit long-distance dependencies between the input sequence, whereas convolutional networks tend to be faster to train and execute. However, the trade-offs between each in terms of the classification accuracy and execution speed are beyond the scope of this paper, which rather focuses on the performance of each network on different hardware.

In each case, the convolutional or recurrent layers operate over an input of shape $[n_{\rm batch}, n_{\rm tracks}, n_{\rm features}]$, where $n_{\rm batch}$ is the number of examples per training or inference batch, $n_{\rm tracks}$ is the number of input tracks, each with $n_{\rm features}$ features. Here, the recurrent network implementation uses a `long short-term memory unit' (LSTM)~\cite{hochreiter1997long}
followed by a number of fully connected layers operating on the output of the last element in the sequence. For the convolutional network, several one-dimensional convolution operations with learnable kernel parameters, are applied sequentially. These convolutional layers are followed by a downsampling `max-pooling' operation that propagates only the maximum of its inputs over a fixed range, and subsequently flattened to one dimension before entering a set of fully connected layers. The corresponding network configuration, and example parameters, can be seen in Table~\ref{taggingNetworkTable}.

\begin{figure}[h!]
\centering
\includegraphics[width=1.0\textwidth,keepaspectratio]{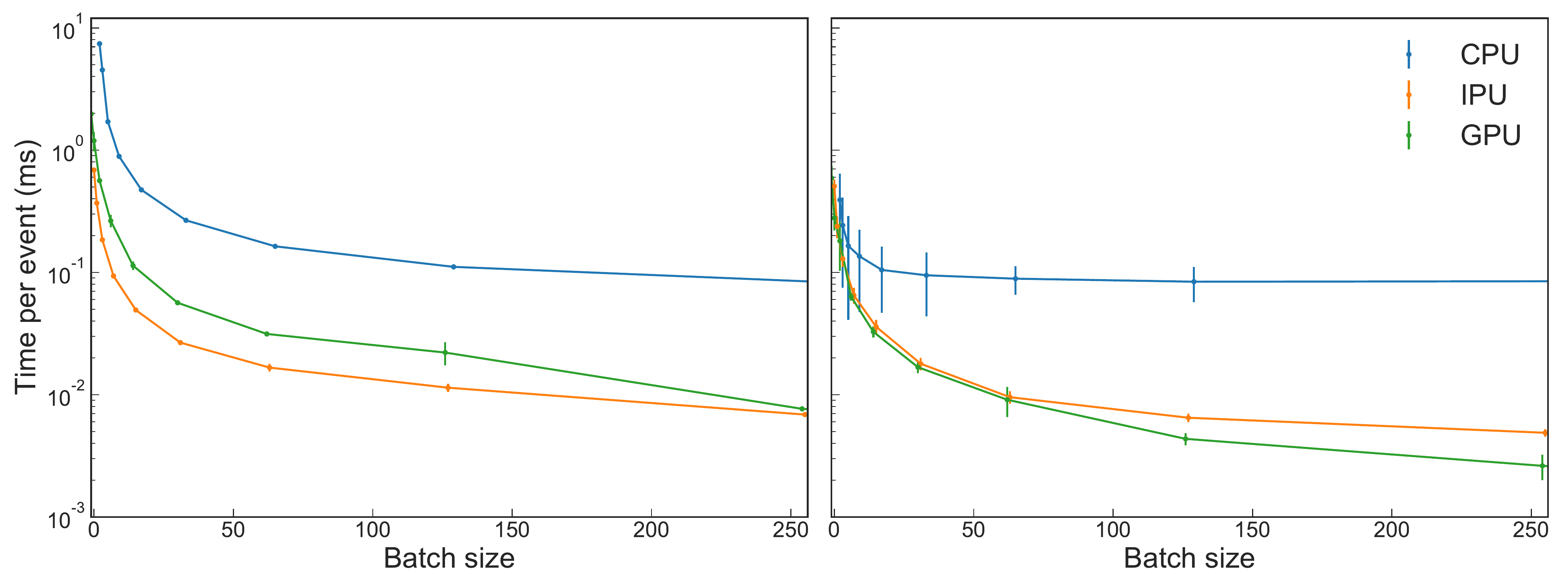}
\caption{Recurrent (left) and convolutional (right) neural network  execution time per event as a function of the batch size.}
\label{tagBatchNN}
\end{figure}

Both of these networks are constructed in PyTorch \texttt{1.2.0}~\cite{paszke2019pytorch}, and exported to the ONNX~\cite{bai2019} interchange format. For execution on the IPU, the ONNX models are imported into the \gc{} PopART framework. For the CPU and GPU benchmarks however, the networks are executed directly in PyTorch, which for GPU execution ensures that the optimised \nvidia{} CuDNN LSTM~\cite{chetlur2014cudnn} implementation is used. The CPU is one single core of an Intel \xeon{} Platinum 8168 processor, the GPU is an \nvidia{} P100 (using \cuda{} toolkit \texttt{10.0} and CuDNN \texttt{10.1}), and the IPU is a \gc{} C2 IPU (using Poplar \texttt{1.0.172}). 

The networks are configured with hyperparameters that result in a modest total number of trainable parameters, whilst still permitting execution in reasonable time for particle physics applications.
A critical parameter that affects inference time, particularly for SIMD processors such as GPUs, is the batch size (\emph{i.e.}, the number of inputs present on the device and executed over in a single inference step).
The variation of inference time per event as a function of the total number of events per batch, can be seen in Figure~\ref{tagBatchNN}. Here, events of size of $n_{\text{tracks}} = 100$ and $n_{\text{features}} = 18$ are used (in addition to the parameters given in Table~\ref{taggingNetworkTable}), which are typical for tagging at LHCb.
As expected, the CPU performance is dominated in all cases by the GPU and IPU, and saturates at large batch size.
For the GPU and IPU, the performance increases as a function of batch size, until the IPU memory capacity is exceeded at a batch size in excess of approximately $256$.
Performances for the convolutional network are approximately equivalent for the GPU and IPU as a function of the batch size, and for the recurrent network, the IPU is a constant $\sim2$ times faster than the GPU at intermediate batch sizes, becoming roughly equivalent at higher batch sizes.
These benchmarks do not consider the time spent copying input and output data across the PCI-e interface.
\begin{table}
\caption{Convolutional and recurrent neural networks used in the flavour tagging example. Parameters correspond specifically to plots in Figure~\ref{tagBatchNN}, and inputs are processed sequentially from the upper to the lower layers, with an implicit sigmoid activation at the end to express the probability of being a $B^0$ or $\overline{B}^0$.}
\begin{center}
\begin{tabular}{ l l }
Convolutional network & Recurrent network \\
\hline
 Conv1D(hidden = 8, k = 20) & LSTM(hidden = 8)  \\
 Conv1D( hidden = 8, k = 10) & Linear(hidden = 8)  \\
 MaxPool1D(pool = 2) &  \\
 Flatten() &  \\
 Linear(hidden = 8) & \\
 Linear(hidden = 8) & \\
\end{tabular}
\label{taggingNetworkTable}
\end{center}
\end{table}

\begin{figure}[h!]
\centering
\includegraphics[width=0.49\textwidth,keepaspectratio]{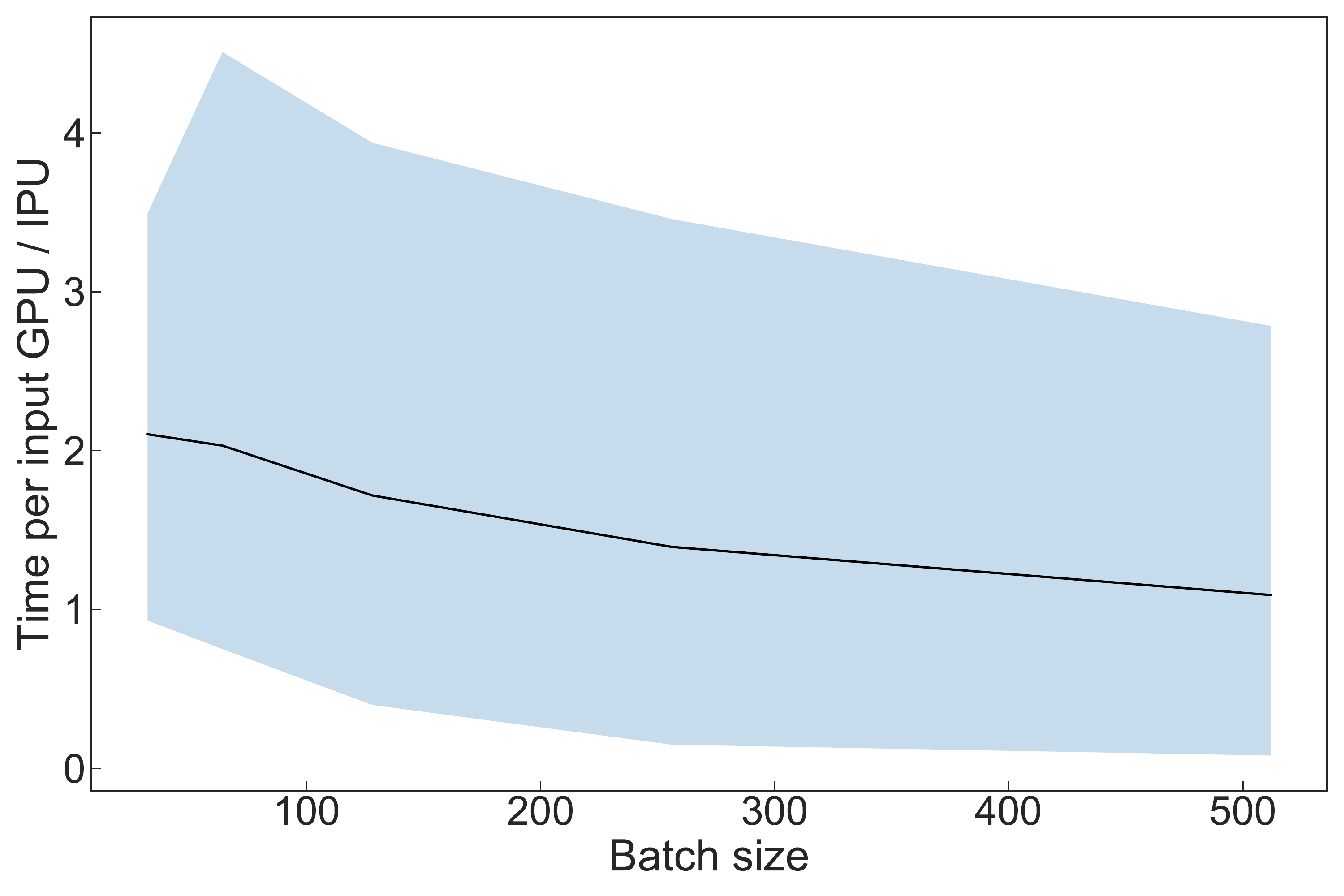}
\includegraphics[width=0.49\textwidth,keepaspectratio]{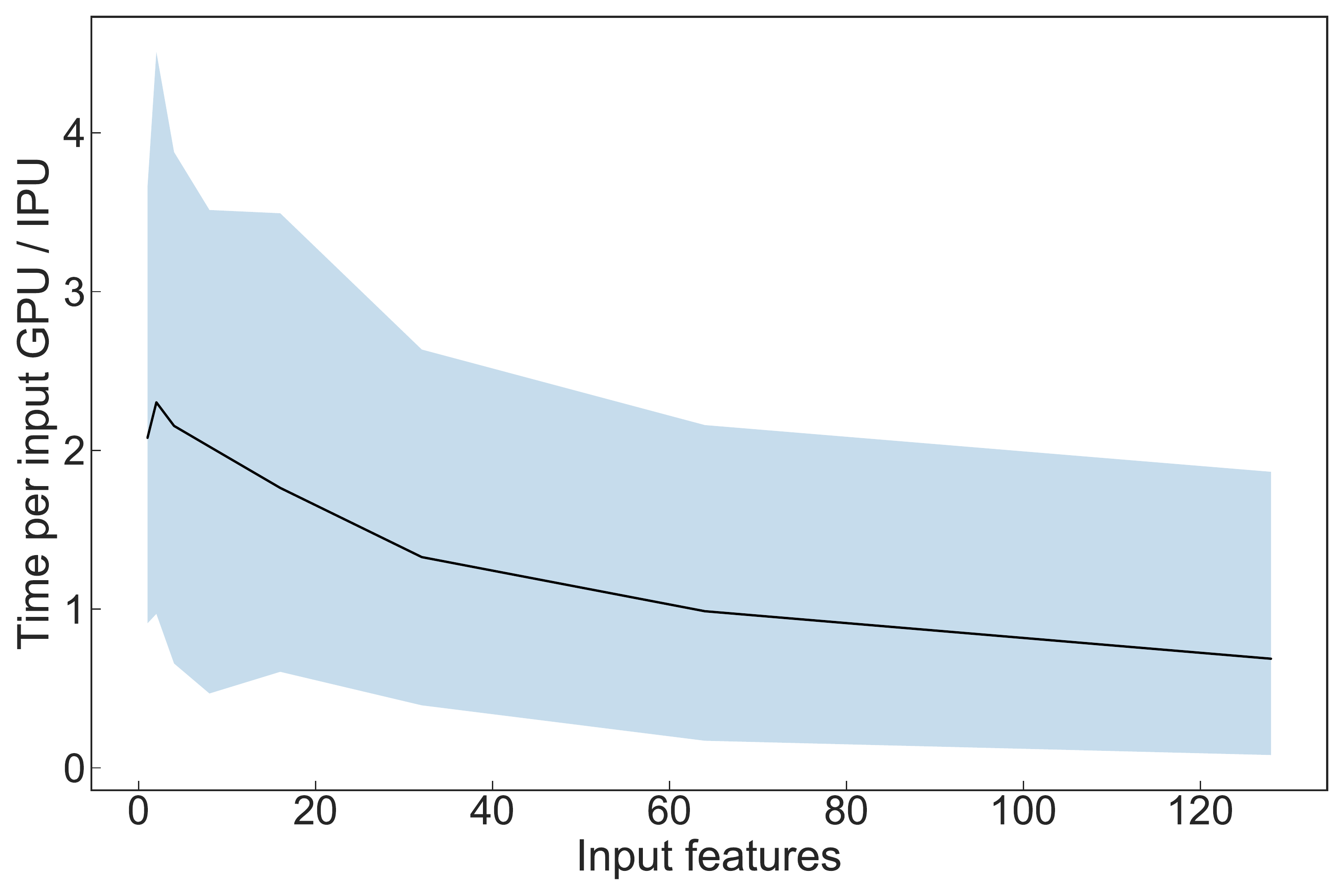}\\
\includegraphics[width=0.49\textwidth,keepaspectratio]{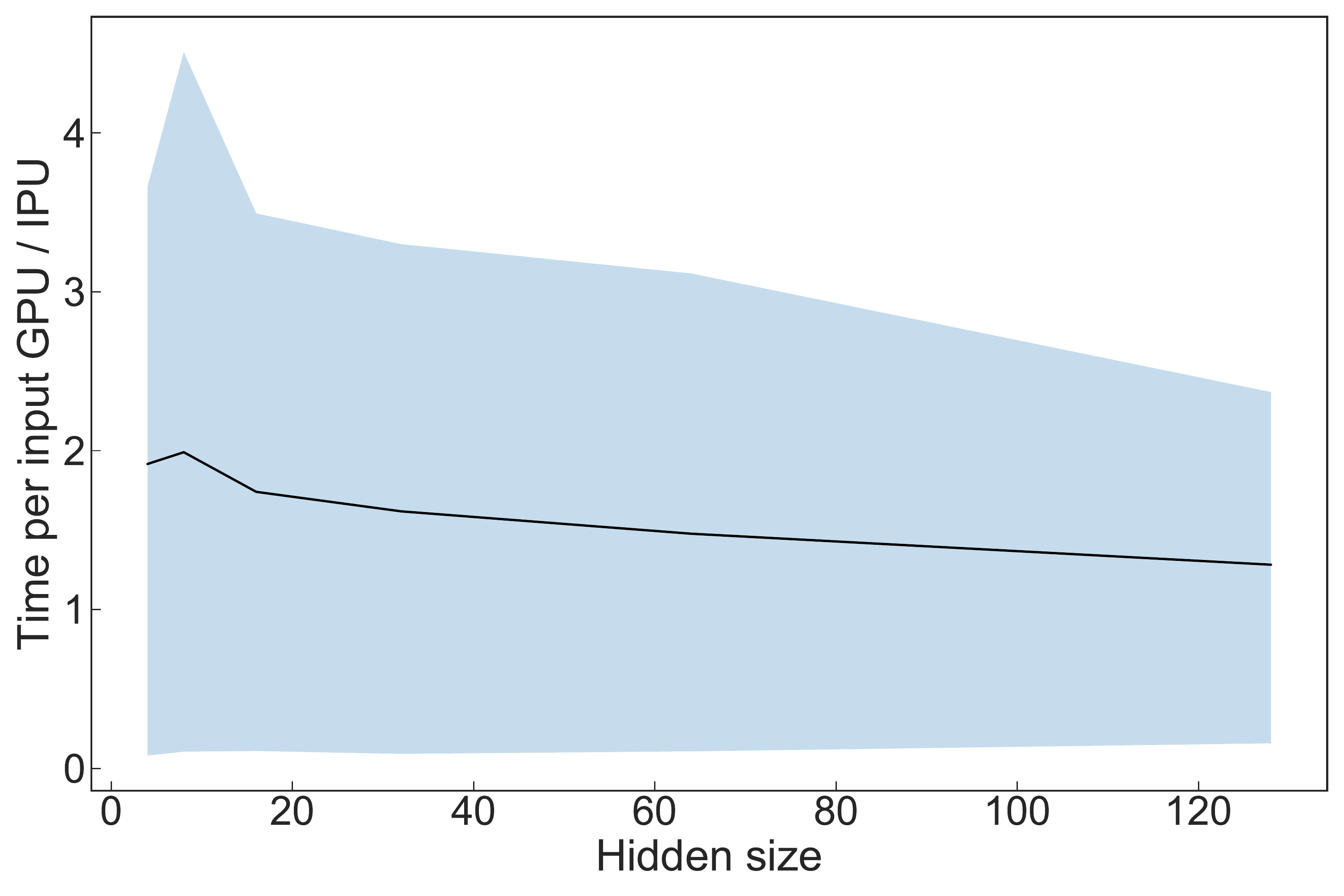}
\includegraphics[width=0.49\textwidth,keepaspectratio]{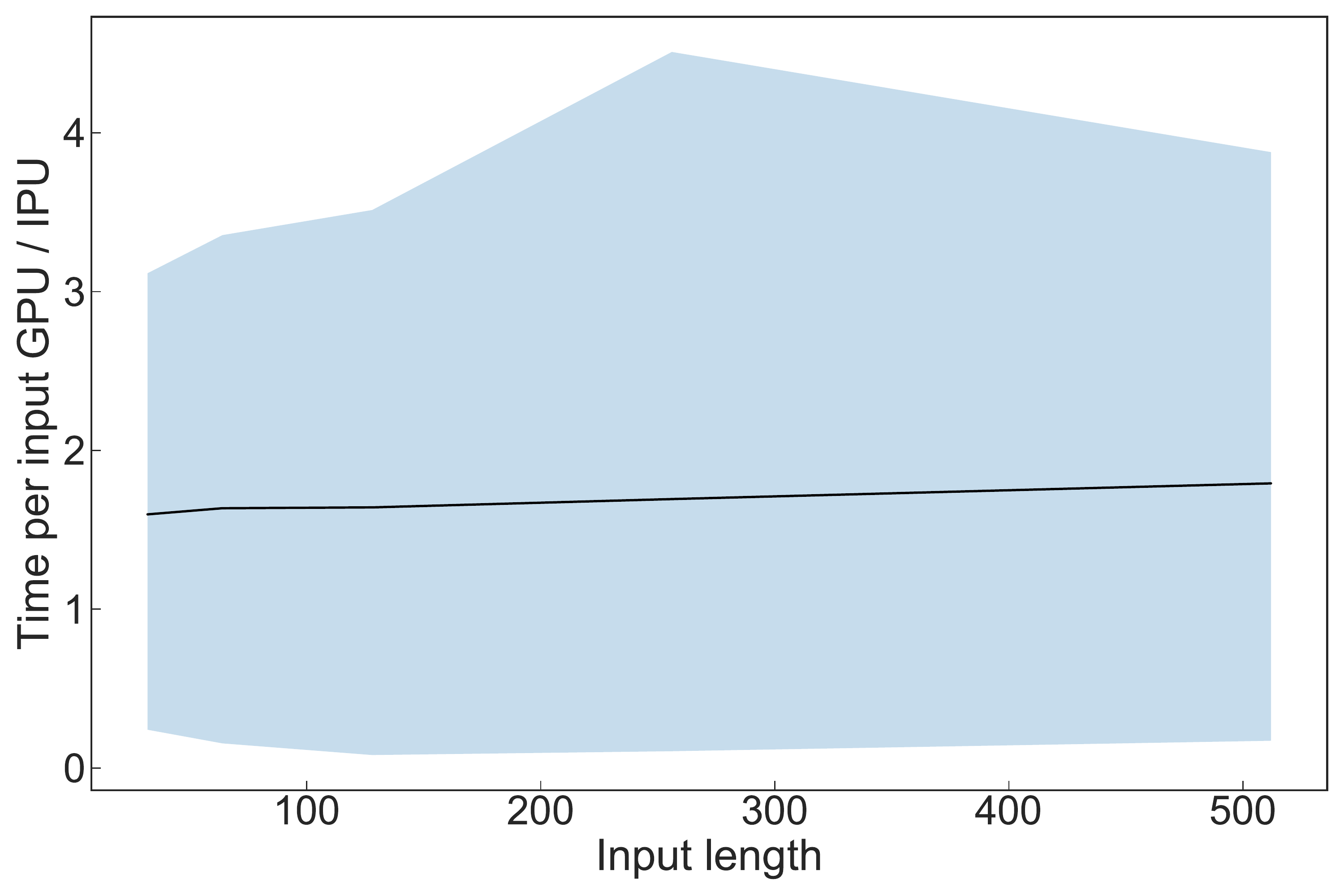}
\caption{Variation of the ratio between the time taken for each input event as a function of batch size, number of input features, hidden layer size, and input length, for the recurrent neural network. In each case, the black curve indicates the average time ratio when holding the $x$-axis value constant, and the coloured band spans the spans the range of possible ratios with constant $x$-axis value.} 
\label{tagRNNScans}
\end{figure}

\begin{figure}[h!]
\centering
\includegraphics[width=0.49\textwidth,keepaspectratio]{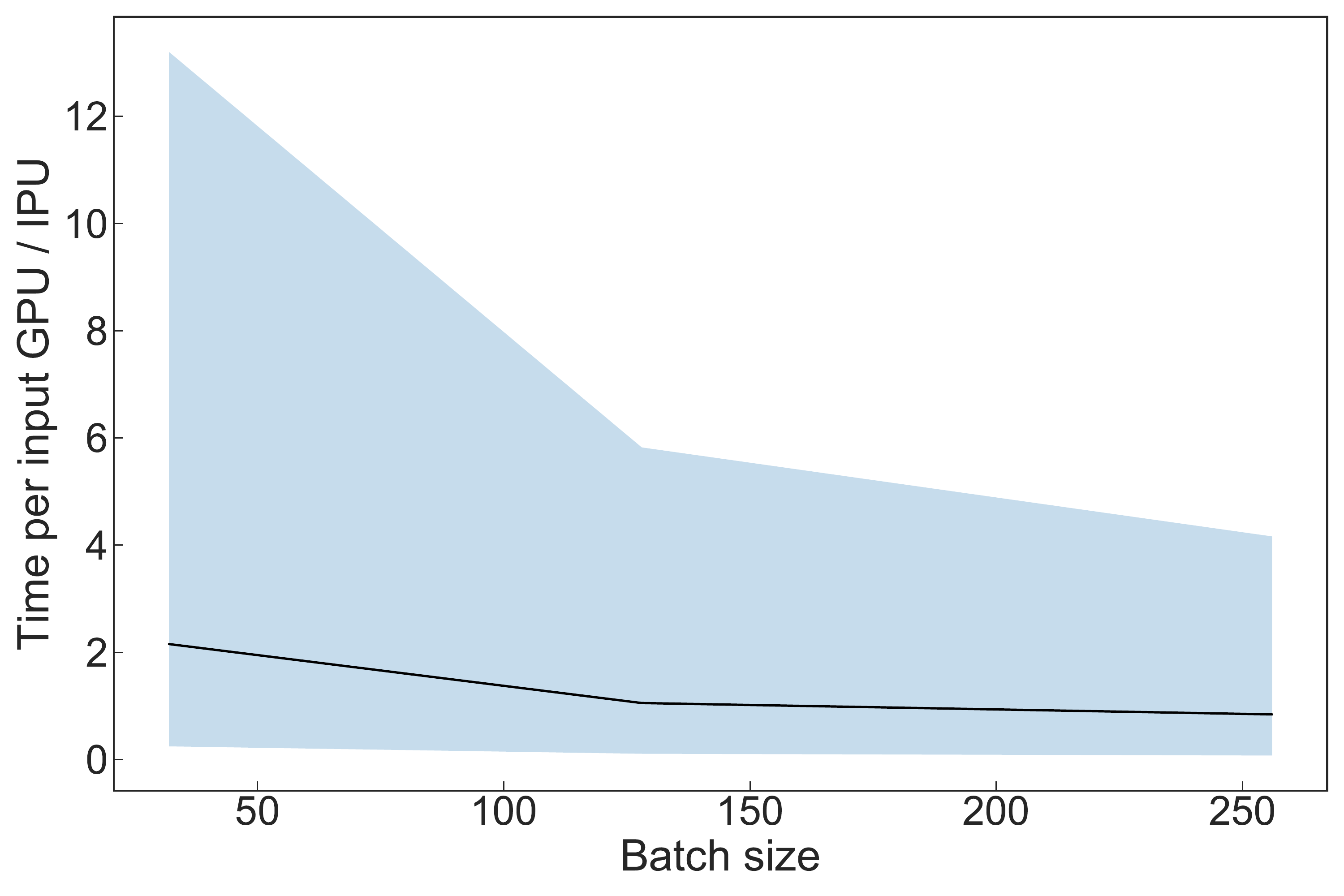}
\includegraphics[width=0.49\textwidth,keepaspectratio]{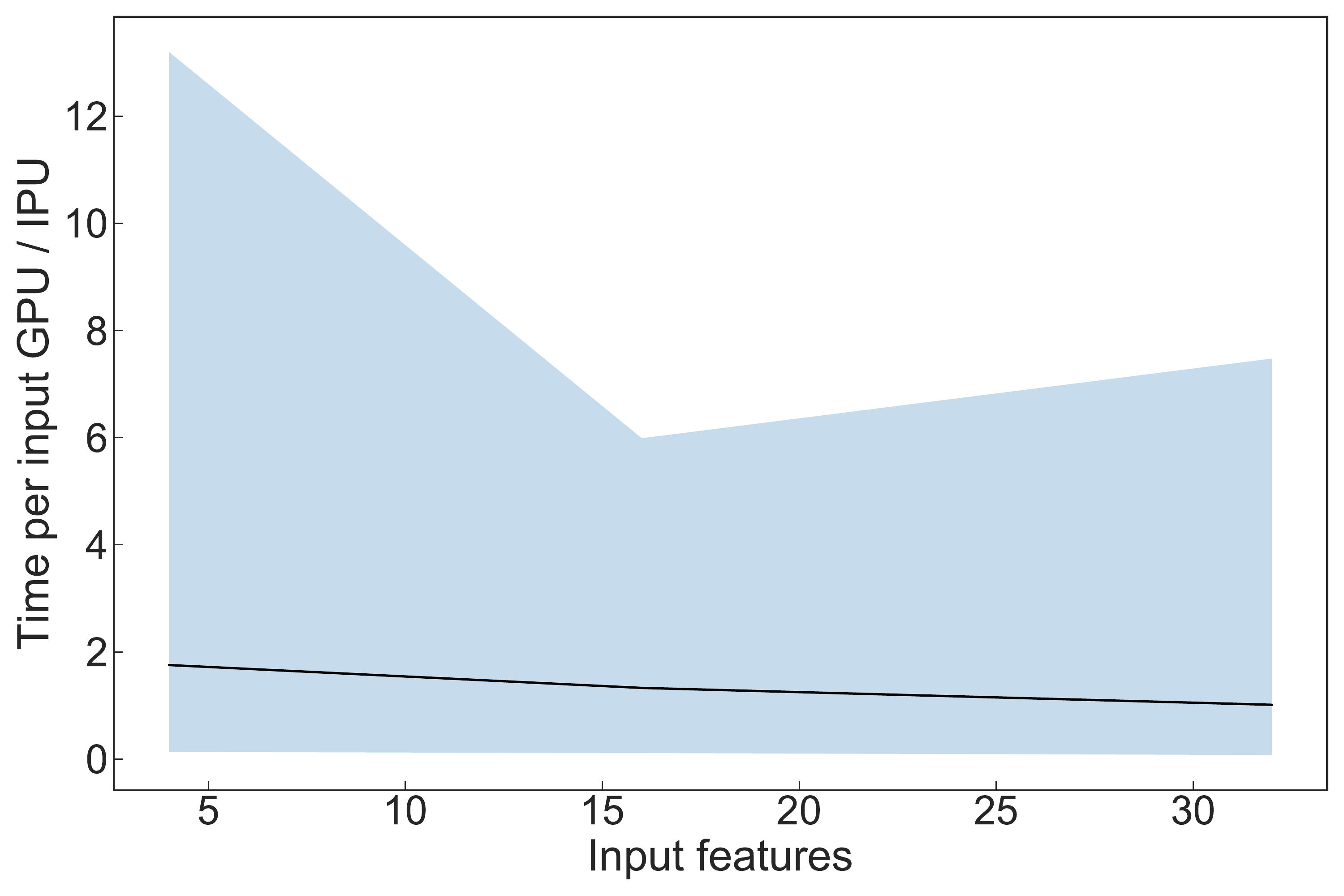}\\
\includegraphics[width=0.49\textwidth,keepaspectratio]{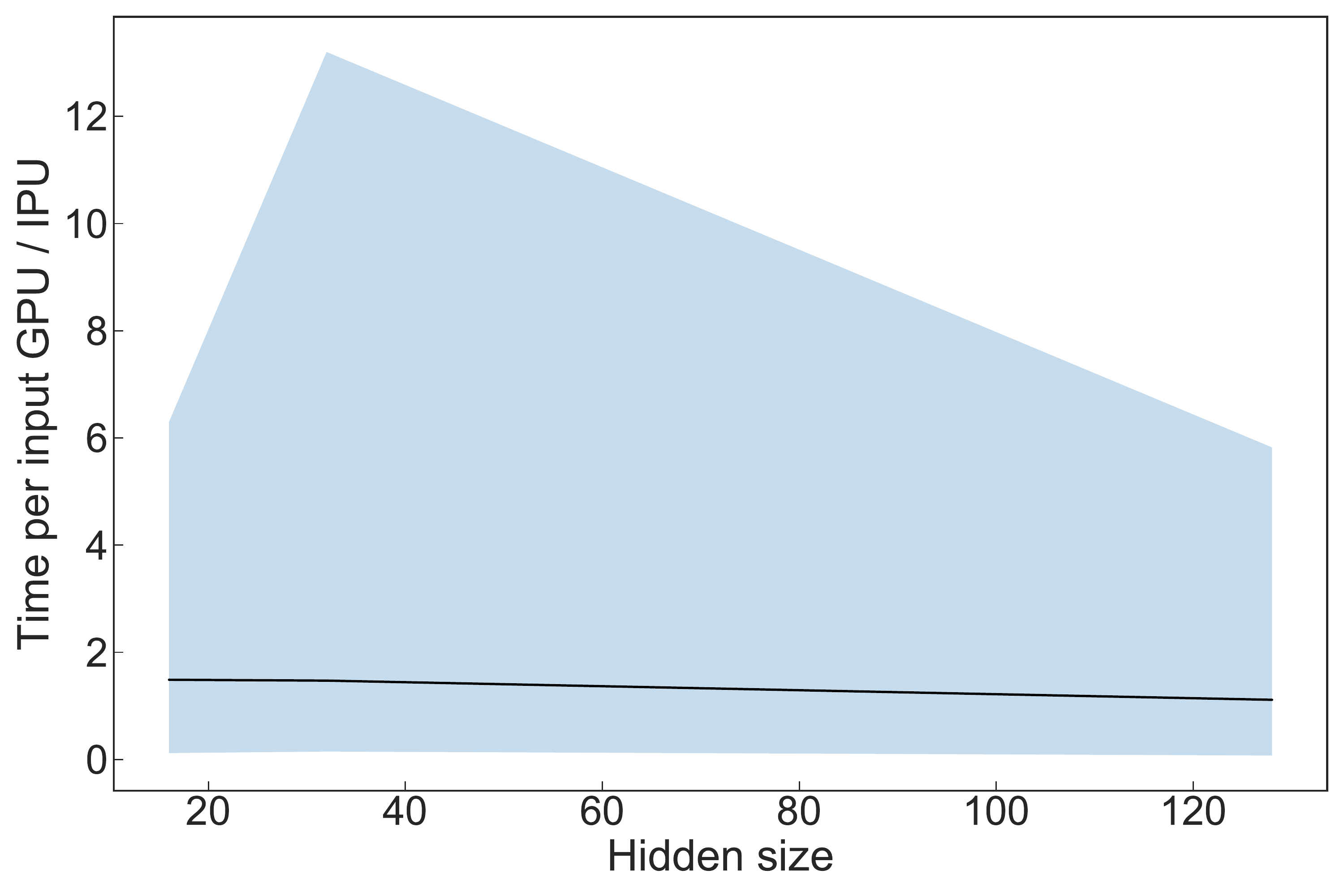}
\includegraphics[width=0.49\textwidth,keepaspectratio]{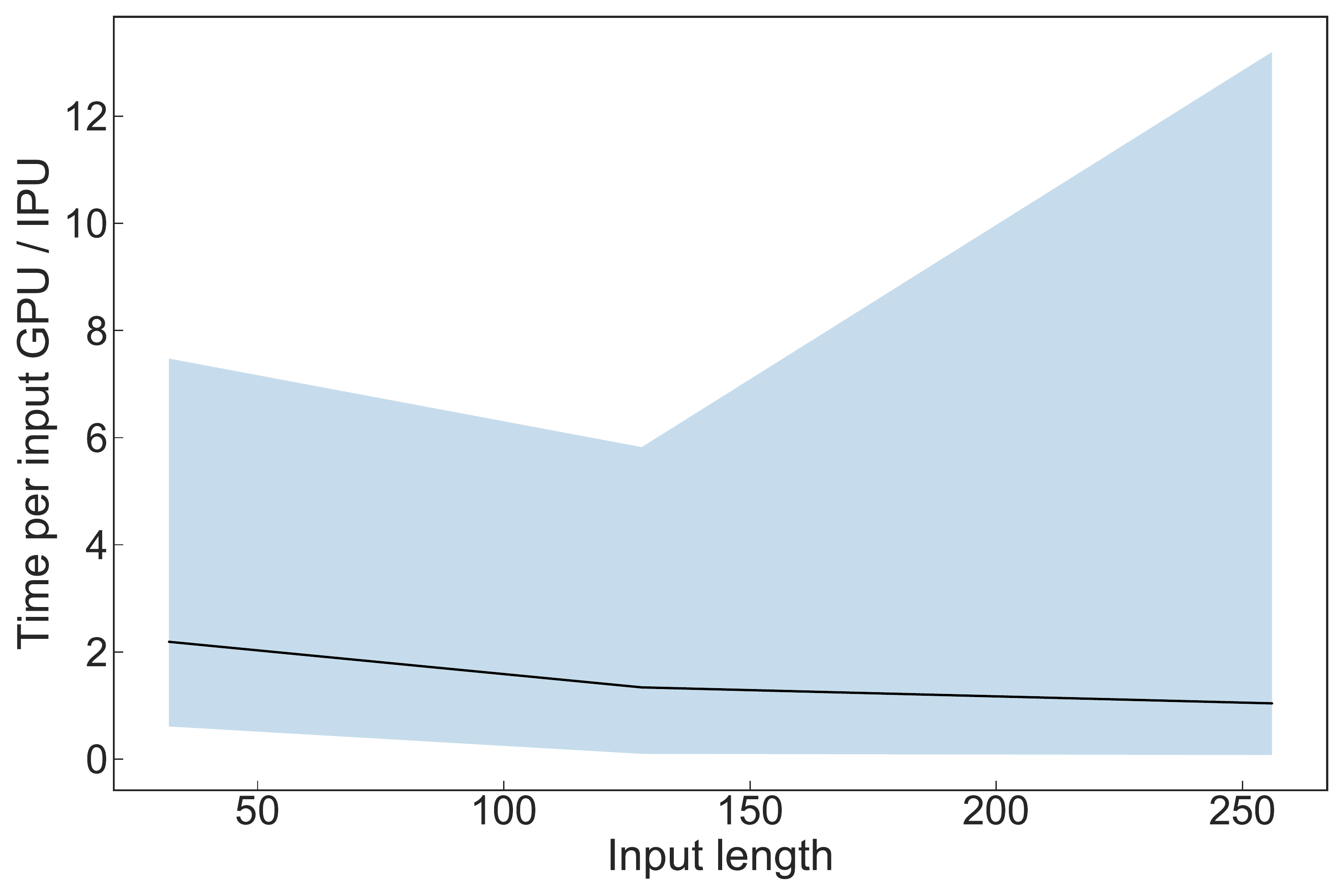}\\
\includegraphics[width=0.49\textwidth,keepaspectratio]{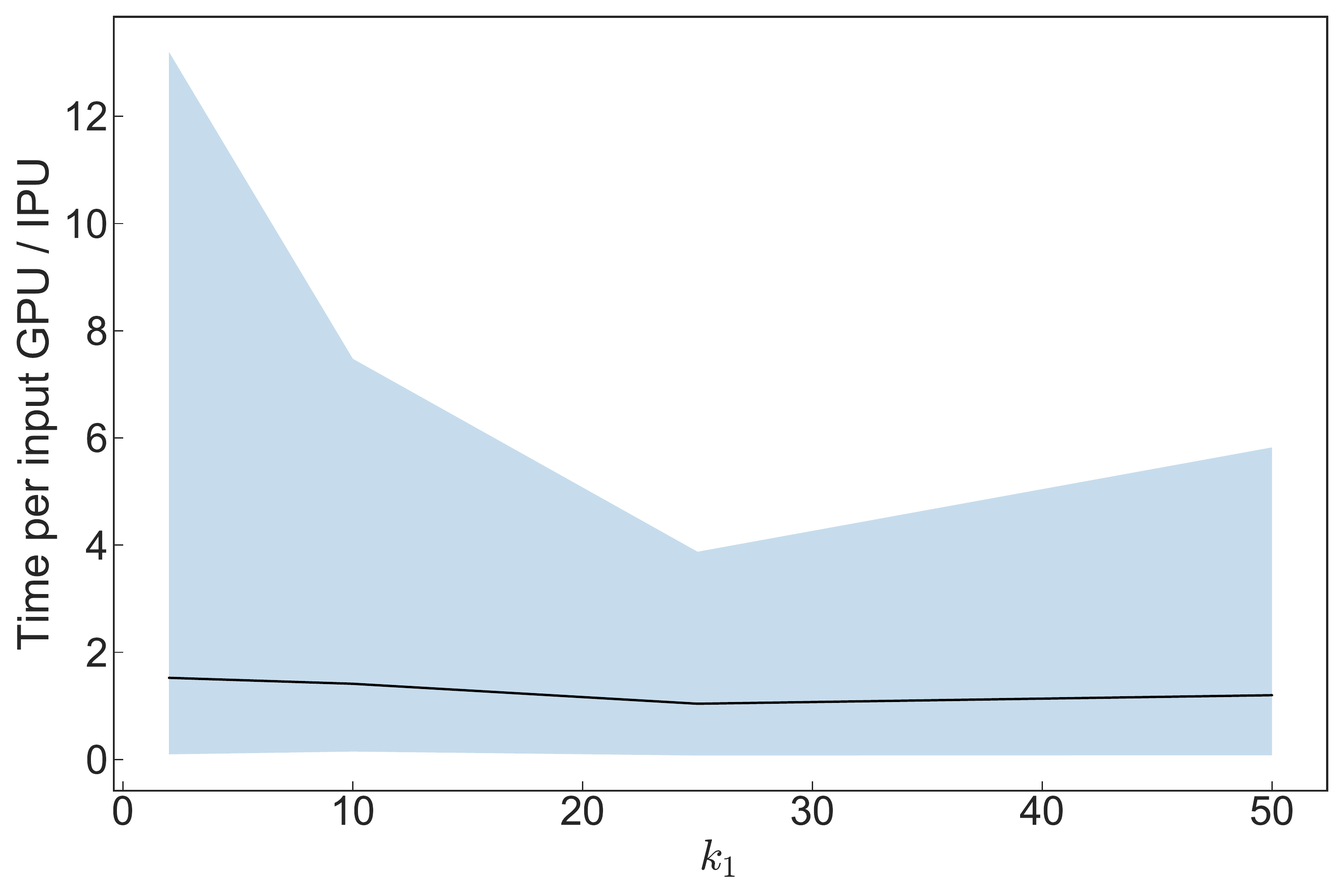}
\includegraphics[width=0.49\textwidth,keepaspectratio]{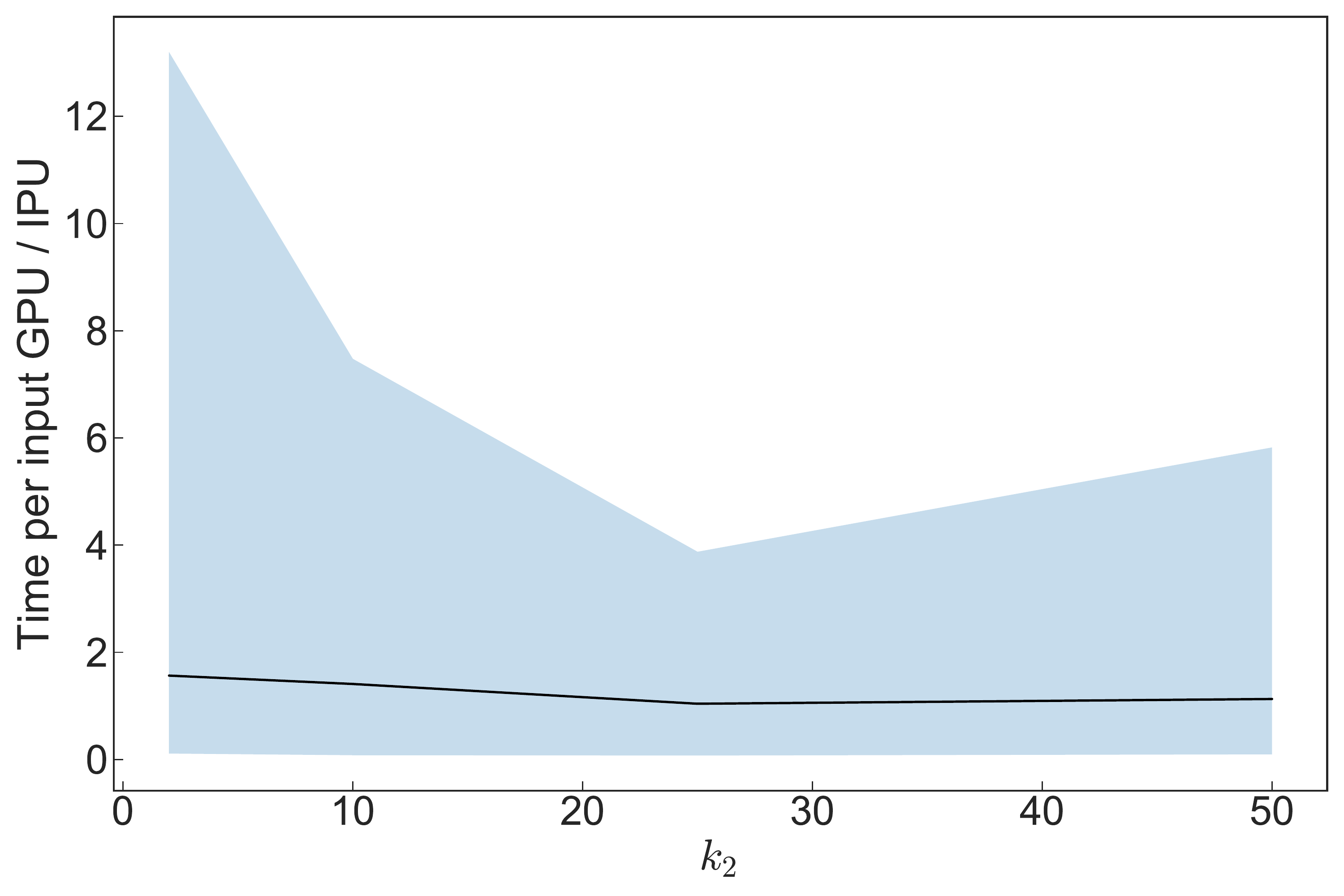}
\caption{Variation of the ratio between the time taken for each input event as a function of batch size, number of input features, hidden layer size, input length, and the size of the two convolutional kernels, for the convolutional neural network. In each case, the black curve indicates the average time ratio when holding the $x$-axis value constant, and the coloured band spans the range of possible ratios with constant $x$-axis value.}
\label{tagCNNScans}
\end{figure}

The batch size is expected to be the dominant factor controlling performance for SIMD processors, all else being equal. However, it is instructive to explore how the variation of network parameters affects relative GPU and IPU performance, particularly given that the IPU does not primarily gain its performance from SIMD processing, so whilst being used for similar purposes, GPUs and IPUs are architecturally quite different.
For the recurrent network architecture, scans are performed over the batch size, number of hidden units (common to each layer), the number of input features per track, and the number of input tracks. Projections of the ratio of the time per input for the GPU and IPU versus each of these parameters can be seen in Figures~\ref{tagRNNScans} and ~\ref{tagCNNScans}.

In each plot, the black curve is the average across all other parameters, holding the $x$-axis parameter constant, and the coloured band spans the minimum and maximum variation of the ratio of execution times.
Therefore, it is expected that if the dependence on relative performance is due to a single of these parameters, then the extent of the coloured band in the plot of this parameter will be small, indicating no or little variation due to the other parameters; at the same time, the black curves in the plots of the other parameters will have little variation as a function of that parameter.

For the RNN in these configurations, we observe a weak dependence on the input length and hidden size, however moderate dependence is seen on the batch size and the number of input features. That no parameter is sufficient to entirely describe the behaviour indicates that the relative performance of the GPU and IPU is a complicated function of all neural network parameters. However, it is clear from these plots that the IPU is better performing for smaller batch sizes, and a smaller number of input features, compared to the GPU.

For the CNN, a more mixed picture is observed, where no single parameter significantly represents the difference between the IPU and GPU performance, however the largest dependence is on the batch size and number of input features. In this case, it is clear that the kernel size has a significant impact on the difference in execution time between the IPU ad GPU, where the IPU tends to perform better in some cases with large values, and in some cases with small values.

\section{K\'{a}lm\'{a}n filter implementations across several architectures}
\label{sec:DansKalmanFilter}
K\'{a}lm\'{a}n filters are a ubiquitous technique for state-space estimation from multiple noisy measurements, and are used in fields as diverse as robotics, signal processing, and econometrics. In particle physics they are most commonly used as a method to incorporate kinematical constraints and detector-material interactions when estimating the particle track state from clustered hits in tracking stations. As such, K\'{a}lm\'{a}n filters often form the basis of event reconstruction algorithms.

Recent emphasis on complete online processing of full events motivates the need for more efficient reconstruction algorithms. In particular, from Run 3 of the LHC, the LHCb experiment intends to perform full event reconstruction at $30$MHz in the high-level trigger, in order to exploit the efficiency gain from performing analysis-level selections earlier in the pipeline. As such, the execution speed of this reconstruction, of which the K\'{a}lm\'{a}n filter is a dominant contributor~\cite{CamporaPerez:2292435}, is strictly limited from a cost-performance perspective.

As many of these operations are inherently parallelisable, implementation of the reconstruction and track filtering on graphics processing units (GPUs) shows good promise, and is potentially a more cost effective alternative to CPUs. Nevertheless, as GPUs are generally designed as single-instruction multiple-data processors, they lack many features that are found in CPUs, such as support for conditional program flow, large caches, and fast interconnects between the compute cores.

\subsection{K\'{a}lm\'{a}n filter formalism}

\label{kalmanFilterFormalism}

\begin{figure}[h!]
\centering
\includegraphics[width=0.60\textwidth,keepaspectratio]{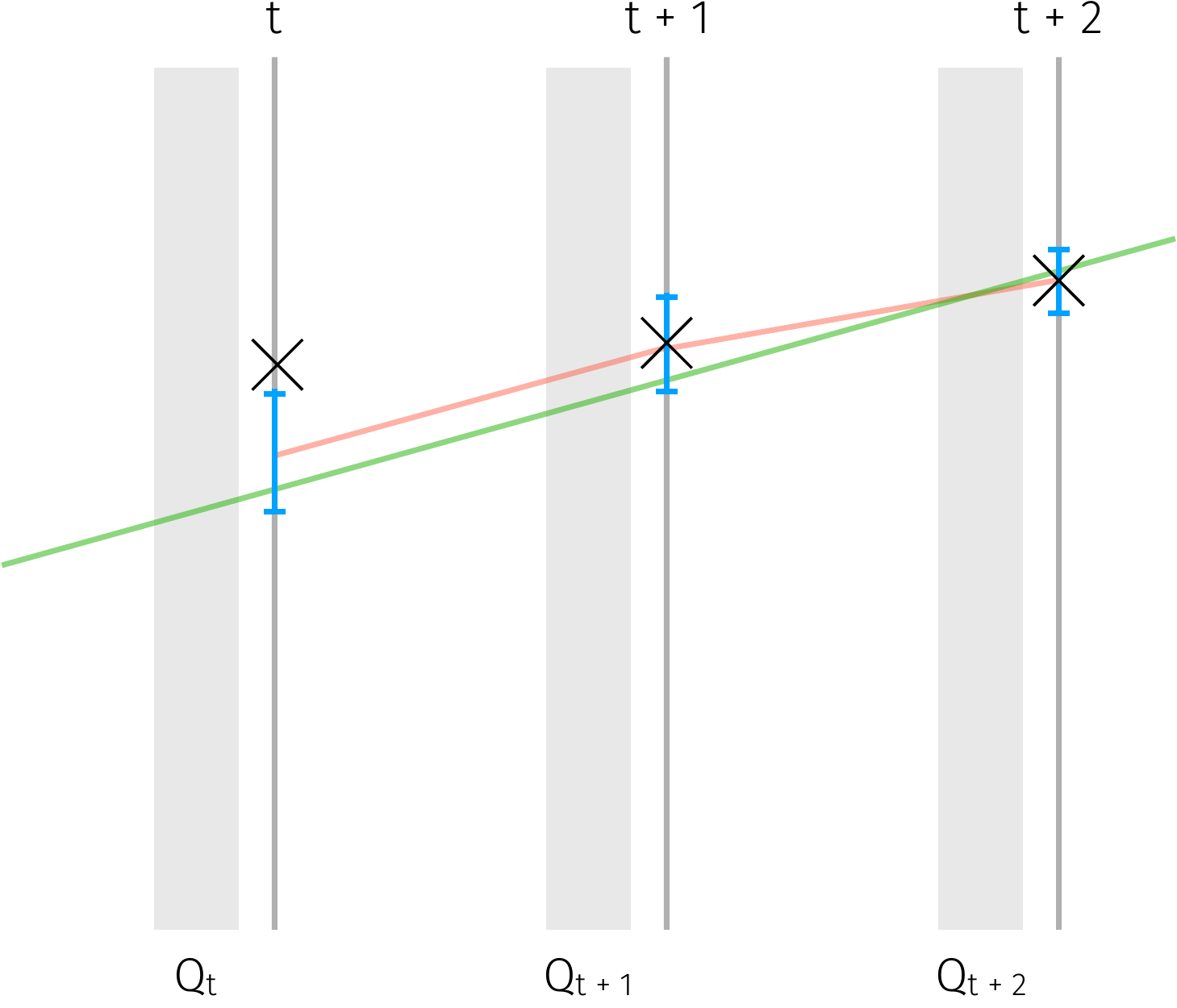}
\caption{Schematic of the K\'{a}lm\'{a}n filter application with active detector planes (dark grey) with hits (crosses), and inactive medium (light grey). The K\'{a}lm\'{a}n filter first calculates the extrapolation of the track state and uncertainty to the next detector plane (blue regions), and corrects this using the true hits and their uncertainties to form an estimate of the track state at the plane (red curve). Lastly, the most likely values of the track states and uncertainties at the planes are obtained in a backwards pass (green curve).}
\label{kalmanDiagram}
\end{figure}

K\'{a}lm\'{a}n filters recursively compute closed-form least-squares estimates for the state and its covariance matrix, under the assumption that all uncertainties can be well described by multidimensional normal distributions; and that only linear relations exist between the state at step $t$ and the state at step $t + 1$, and the state and the measurement process. The application of a K\'{a}lm\'{a}n filter can be broken down into three stages: a prediction (or projection) stage where the state at step $t$ is projected linearly to a state at step $t + 1$; a filtering stage where the state at step $t + 1$ is corrected using the measurement and covariance matrix of the measurement at step $t + 1$; and a smoothing stage after all filtering steps, where state and covariance matrix updates are propagated backwards through the states to achieve a globally optimal configuration. The formulation here follows that of Refs.~\cite{Fruhwirth:1987fm, hernando1998kalman}.

The first projection step is described by a set of recurrence relations that extrapolate the state described by a vector $\vec{p}$
at step $t$ to the values at step $t + 1$, given by
\begin{equation}
    \vec{p}_{t + 1, \text{proj}} = \mathbf{F}_t \vec{p}_t,
    \label{eq:projection}
\end{equation}
with the covariance matrix of $\vec{p}$ given by $\mathbf{C}$, where
\begin{equation}
    \mathbf{C}_{t + 1, \text{proj}} = \mathbf{F}_t \mathbf{C}_t \mathbf{F}_t^{\top} + \mathbf{Q}_t.
    \label{eq:covProjection}
\end{equation}
 These relations are expressed in terms of the transfer matrix $\mathbf{F}_t$, and the random error matrix $\mathbf{Q}_t$. The expression in Eq.~\ref{eq:projection} uses the underlying modelling assumptions (in the case of this particular track reconstruction, simple kinematics) that generate $p_{t + 1}$ from $p_t$ via the application of the linear operator $\mathbf{F}_t$. The error matrix $\mathbf{Q}$ contains the process noise that involves terms that describe additive errors to the estimated state, such as those that are picked up after each propagation step from material interactions, \emph{etc}.

At step $t + 1$, the prediction from step $t$ to $t + 1$, $\vec{p}_{t + 1, \text{proj}}$ is updated using the measurements at $t + 1$, $\vec{m}_{t + 1}$. The relation between the measurement $\vec{m}$ and the state $\vec{p}$ is given by $\mathbf{H}$ (which in general is independent of $t$), and the updated \emph{filtered} expectation of $\vec{p}_{t + 1}$ becomes

\begin{equation}
    \vec{p}_{t + 1, \text{filt}} = \mathbf{C}_{k + 1, \text{filt}} \left[ \mathbf{C}^{-1}_{t + 1, \text{proj}} \vec{p}_{t + 1, \text{proj}} + \mathbf{H}^{\top} \mathbf{G}_{t + 1} \vec{m}_{t + 1} \right],
    \label{eq:filtering}
\end{equation}
where
\begin{equation}
    \mathbf{C}_{t + 1, \text{filt}} = \left[ \mathbf{C}_{t + 1, \text{proj}} + \mathbf{H}^{\top} \mathbf{G}_{t + 1} \mathbf{H} \right].
    \label{eq:covFiltering}
\end{equation}
is the corresponding covariance matrix. Here, $\mathbf{G}_t$ is the matrix that describes weights corresponding measurement noise, such as the detector resolution, at step $t$.

Up until this point, all information is updated in the forward direction, however information downstream can also be used to update upstream state estimates, in order to obtain a globally optimal set of states. To do this propagation, a backward transport operator is defined as
\begin{equation}
    \mathbf{A}_t = \mathbf{C}_{t + 1, \text{filt}} \mathbf{F}_t^{\top} \mathbf{C}^{-1}_{t + 1, \text{proj}},
    \label{eq:backTransport}
\end{equation}
which is used to perform the \emph{smoothing} step in the backward direction and updating the state
\begin{equation}
    \vec{p}_{t, \text{smooth}} = \vec{p}_{t, \text{filt}} + \mathbf{A}_t( \vec{p}_{t + 1, \text{smooth}} - \vec{p}_{t + 1, \text{proj}}),
    \label{eq:smooth}
\end{equation}
and covariance matrix
\begin{equation}
    \mathbf{C}_{t, \text{smooth}} = \mathbf{C}_{t, \text{filt}} + \mathbf{A}_t( \mathbf{C}_{t + 1, \text{smooth}} - \mathbf{C}_{t + 1, \text{proj}}) \mathbf{A}_t^{\top},
    \label{eq:smoothCov}
\end{equation}
at $t$ using the now smoothed state and covariance matrix at $t + 1$.

The covariance matrix can also be used to form a $\chi^2$ test statistic to determine the consistency of a hit with the fitted track,

\begin{equation}
    \chi^2_t = \vec{r}^T_t \mathbf{G}_t \vec{r}_t + (\vec{p}_{t, \text{filt}} -p_{t, \text{proj}}) \mathbf{C}_{t,\text{proj}}^{-1} (\vec{p}_{t, \text{filt}} -p_{t, \text{proj}}),
    \label{eq:kalmanChiSq}
\end{equation}
where $r_k$ is the residual,
\begin{equation}
    \vec{r}_k = \vec{m} - \mathbf{H} \vec{p}_{t, \text{filt}}.
\end{equation}

\subsection{K\'{a}lm\'{a}n filter configuration}

\label{kalmanFilterConfig}

To investigate the performance characteristics of a K\'{a}lm\'{a}n filter implemented in Poplar on the IPU, a tracker with 2D active planes of $1$m $\times$ $1$m in $\hat{x}-\hat{y}$ is considered, separated by a homogeneous inactive medium that induces multiple scattering. Five of these planes are used, separated in $\hat{z}$ by $d = 1$m of the inactive medium, and indexed by $t$. Each of these detector planes record measured track hits, $\vec{m} = \{m_x, m_y\}$, discretised according to the physical resolution of the detector planes, $\sigma$.

No magnetic field is considered, however its inclusion would only result in a minor modification of the track state (to infer momentum) and inclusion of the magnetic field description in $\mathbf{F}.$ It is assumed initially that each track registers a hit on each of the five planes, and the matching of hits to tracks is perfect. In reality, dummy hits can be introduced to the tracking algorithms, and tracks are often post-processed to find the most likely set, so neither of these effects compromise the generality of this proof of principle.

\begin{figure}[h!]
\centering
\includegraphics[width=0.80\textwidth,keepaspectratio]{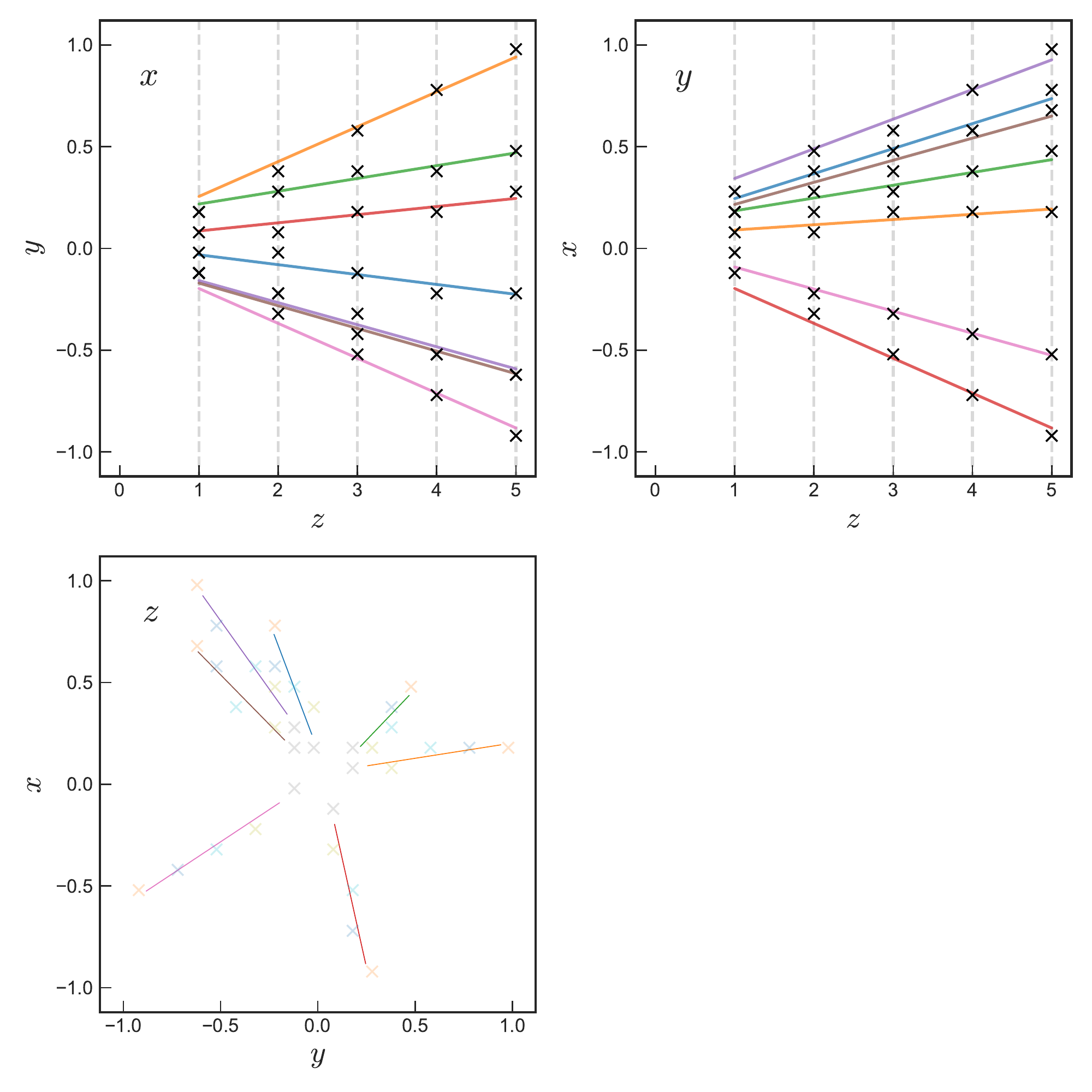}
\caption{Projections of the tracks (coloured lines) reconstructed from hits (crosses) using the detector and K\'{a}lm\'{a}n filter configuration given in the text.}
\label{kalmanRecoPlot}
\end{figure}

A state vector, $\vec{p}_t = \{x_t, \tan{\theta_t}, y_t, \tan{\phi_t}\}$, corresponding to the most likely values of the track $x$-position, $x_t$; $y$-position, $t_t$; tangent of the track slope in $\hat{x}-\hat{z}$, $\tan{\theta}$; and tangent of the track slope in $\hat{y}-\hat{z}$, $\tan{\phi}$; is estimated at each plane, $t$. It follows that the model parameters for such a system are

\begin{align}
\mathbf{F} &= \begin{bmatrix}
1 & d & 0 & 0 \\
0 & 1 & 0 & 0 \\
0 & 0 & 1 & d \\
0 & 0 & 0 & 1 \\
\end{bmatrix},
\hspace{0.25cm}
\mathbf{G} = \begin{bmatrix}
1/\sigma^2 & 0 & 0 & 0 \\
0 & 0 & 0 & 0 \\
0 & 0 & 1/\sigma^2 & 0 \\
0 & 0 & 0 & 0 \\
\end{bmatrix},\\[4ex]
\hspace{0.25cm}
\mathbf{H} &= \begin{bmatrix}
1 & 0 & 0 & 0 \\
0 & 0 & 0 & 0 \\
0 & 0 & 1 & 0 \\
0 & 0 & 0 & 0 \\
\end{bmatrix},
\hspace{0.25cm}
\mathbf{Q} = \begin{bmatrix}
z_0^2\theta_0^2 & z_0\theta_0^2 & z_0^2\theta_0^2 & z_0\theta_0^2 \\
z_0\theta_0^2 & \theta_0^2 & z_0\theta_0^2 & \theta_0^2 \\
z_0^2\theta_0^2 & z_0\theta_0^2 & z_0^2\theta_0^2 & z_0\theta_0^2 \\
z_0\theta_0^2 & \theta_0^2 & z_0\theta_0^2 & \theta_0^2 \\
\end{bmatrix},
\end{align}
where the parameterisation for $\mathbf{Q}$ is obtained from Ref.~\cite{Wolin:1992ti} disregarding higher order terms in the track slopes; $z_0$ is the material depth; and $\theta_0^2$ is the variance of the multiple scattering angle.

The initial state for the first projection step is set to be equal to the hits on the first plane, $\vec{p}_{0,\text{proj}} = \{ m_{0,x}, 0, m_{0, y}, 0 \}$, and the covariance matrix set to equal the full uncertainty on the track state,
\begin{equation}
\mathbf{C}_{0, \text{proj}} = \begin{bmatrix}
(\Delta x) ^2 & (\Delta \tan \theta) ^2 & 0 & 0 \\
0 & 0 & 0 & 0 \\
0 & 0 & (\Delta y) ^2 & (\Delta \tan \phi) ^2 \\
0 & 0 & 0 & 0 \\
\end{bmatrix}.
\end{equation}
where $\Delta x = \Delta y = 1$m, and $\Delta \theta = \Delta \phi = 1$.

In this study, simulated particles are produced at $(0, 0, 0)$ and travel in the positive $\hat{z}$ direction towards the detector planes. At each plane, the particle interacts with the active detector material according to its projection on the $\hat{x}-\hat{y}$ plane of the detector, with a location that is subject to a random fluctuation in each direction depending on the total path length to simulate the effect of multiple scattering. Subsequently the location of the hit is discretised according to the granularity of the active detector area. These two effects determine the K\'{a}lm\'{a}n-filter process and covariance matrices of the measurement uncertainty. An example of the simulated detector configuration can be seen in Fig.~\ref{kalmanRecoPlot}, with the corresponding hits and reconstructed track states.

\subsection{Benchmarks}

\begin{figure}[h!]
\centering
\includegraphics[width=0.70\textwidth,keepaspectratio]{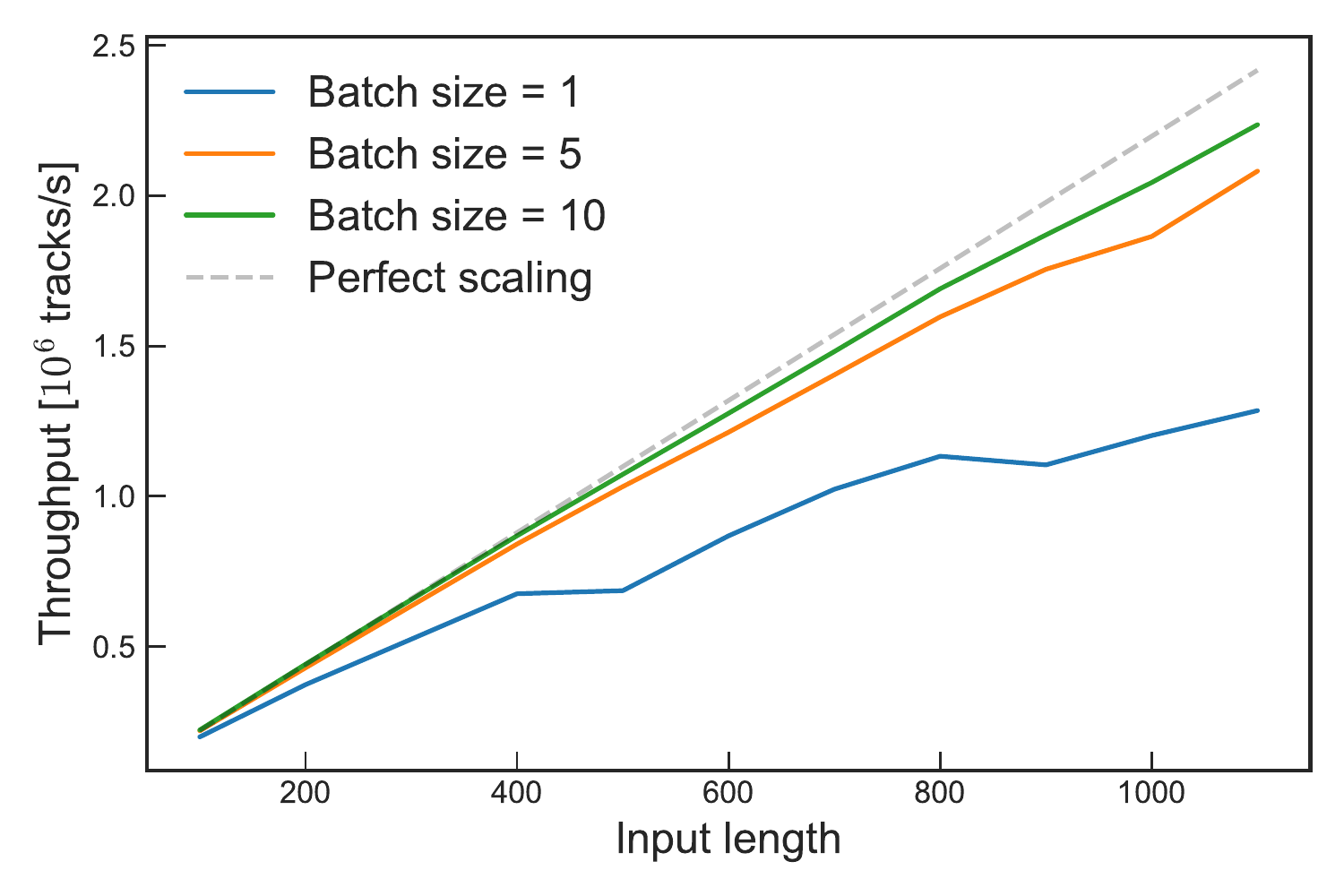}
\caption{Tracks per second processed by the K\'{a}lm\'{a}n filter, as a function of the number of tracks processed in parallel on the tiles (`input size'). This is given for the cases where multiple `batches' of this size are copied to the tiles before execution. The theoretical maximum throughput evolution as a function of input size is also indicated.}
\label{kalmanBatchComparison}
\end{figure}

The K\'{a}lm\'{a}n filter described in Sec.~\ref{kalmanFilterConfig} is implemented for the IPU hardware using the Poplar C++ SDK. To exploit the independence of the particle tracks, each track is assigned to a single IPU tile, where all operations in Sec.~\ref{kalmanFilterFormalism} are performed. In principle this results in $1,216$ K\'{a}lm\'{a}n filter operations proceeding in parallel, however, optimal throughput is only achieved when several batches of tracks are copied to each tile initially, and then operated on sequentially. From Fig.~\ref{kalmanBatchComparison} it can be seen that for batches of size greater than $\sim10$ tracks, almost perfect parallelism is achieved, with a peak throughput of around $2.2\times10^{6}$ tracks per second for this configuration. 

It is interesting to study the behaviour of the IPU implementation of the K\'{a}lm\'{a}n filter with a workload that relies on program branch statements and random memory accesses. To this end, a modification of the above K\'{a}lm\'{a}n filter configuration is implemented, where a proportion of hits are forced to be inconsistent with tracks they have been assigned to. This results in a large value of the $\chi^2$ expression in Eq.~\ref{eq:kalmanChiSq}. At each step the $\chi^2$ value is evaluated, and if it is above a certain threshold, the state is not updated and the previous state is propagated to the next state under the assumption that no hit was observed at this stage.

\begin{figure}[h!]
\centering
\includegraphics[width=0.49\textwidth,keepaspectratio]{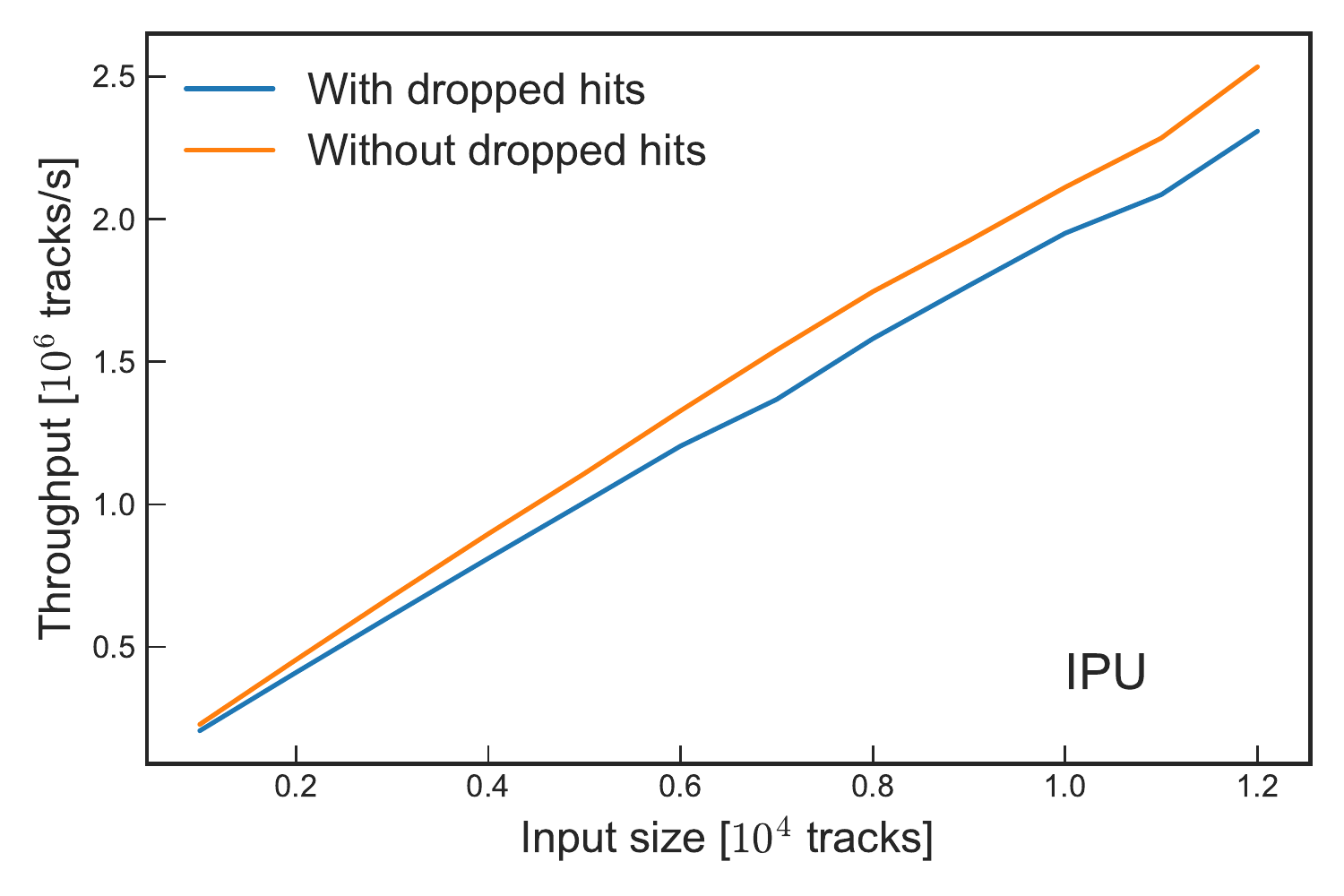}
\includegraphics[width=0.49\textwidth,keepaspectratio]{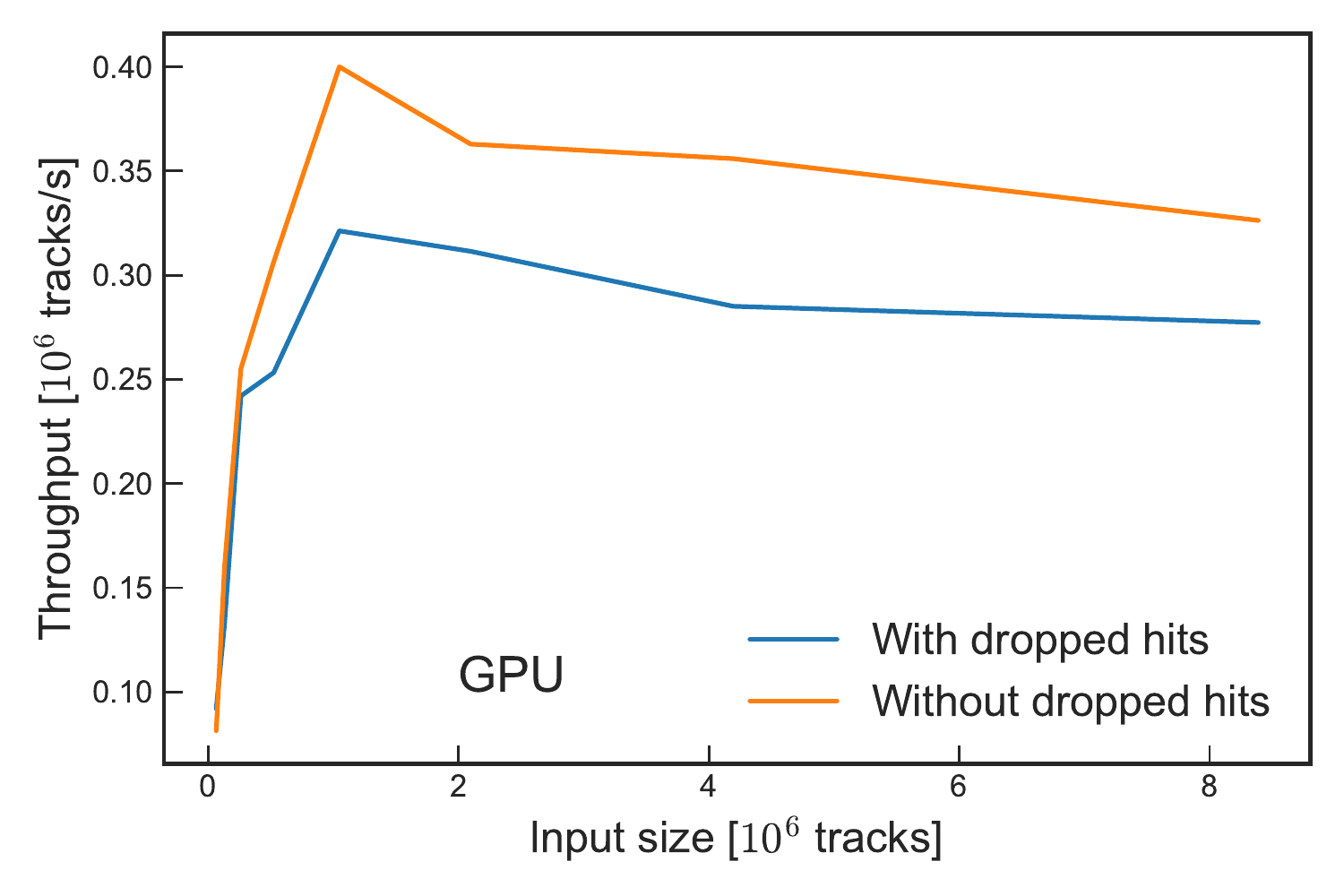}
\caption{Comparison of the throughput with and without conditional execution to ignore mis-associated hits, for a Poplar implementation on the IPU (left), and for a TensorFlow implementation on the GPU (right). The difference in the x-axis scale between the two plots is due to the different memory capacity of the IPU compared to the GPU.}
\label{droppedHits}
\end{figure}

On the IPU, this is implemented by a branch statement in the vertex code, which is executed on each tile separately. By way of comparison, an equivalent K\'{a}lm\'{a}n filter  configuration is also implemented in TensorFlow (\texttt{v2.1.0}) for execution on the GPU.
In TensorFlow the subsequent filtering step is modified using a conditional gather-scatter update to the state and state propagation parameters. Despite the sub-optimal TensorFlow-based GPU implementation, it is instructive to compare the relative throughput in the case where the states are conditionally modified, and the case where no conditional execution is performed. This comparison can be seen in Fig.~\ref{droppedHits}. On the IPU, the reduction in peak throughput is approximately half that of the GPU, likely because the conditional execution results in an inefficiency caused by divergence of parallel threads on the GPU (`warp divergence'), whereas on the IPU these execute independently.

\section{Summary and Conclusions}
\label{sec:conclusion}
This paper represents the first study of IPUs, a new processor type optimised for ML applications, in the context of particle physics. 
\Tf{} and \pytorch{}-based ML applications were used to compare the performance of a 1st generation IPU to that of a GPU of comparable price, but with twice the power consumption, and two high-end CPUs (see \Tabref{tab:hardware}). 
Both GPU and IPU outperform the CPUs. The performance of the IPU and GPU is studied for a variety of neural network architectures and parameters. The batch size is identified as a key variable. 
For batch sizes accessible to both processors, the IPU out-performs the GPU, in some cases by orders of magnitude. For GAN event generation, large batch sizes are usually optimal. Here, the larger memory capacity of the GPU, allowing larger batch sizes, can be a decisive advantage. 
This is the case for the fully-connected GAN architectures studied; for the convolutional- and locally-connected GANs, the IPU generates events faster than the GPU despite using a smaller batch size. It is worth noting in this context that the 2\nd{} generation IPU has triple the memory per tile compared to the 1\ste{} generation IPU used here.
In all cases, GANs train faster on the IPU. For applications with small batch size $\lesssim\mathcal{O}(100)$, such as neural network training or the track-correction algorithm studied, the IPU outperforms the GPU, typically by a factor of 4-5.

This paper also presents the first implementation of a \Kmf{} on an IPU. The algorithm is implemented using \gc{}'s \poplar{} SDK, and re-implemented on a GPU using \Tf{}. While the IPU implementation is much faster, the two implementations are too different for a fair comparison. Comparing the processing speeds on each processor 
with and without the final clean-up step indicates that the IPU's MIMD architecture is a significant advantage when executing conditional control-flow programs.

An important factor in considering the usefulness of IPUs in particle physics, alongside their performance, is the ease with which they can be programmed. The IPU software for the studies presented here~\cite{codeDOI} was written within less than 6 months of the group's first access to \gc{}'s IPUs, by a small team of particle physics postdocs and PhD students with no prior experience of IPU programming.

This first investigation of IPUs in a particle physics context suggests that IPUs, due to a combination of performance, flexibility and ease of programming, have the potential to play a central role in meeting the fast-increasing compute needs of particle physics. As promising as these results are, they can only be a starting point that motivates further, detailed study using realistic particle physics workflows.

\acknowledgments
We are grateful to \gc{} for providing cloud access to their IPUs and for technical support. We also benefited from using the computational facilities of the Advanced Computing Research Centre, University of Bristol - \url{http://www.bris.ac.uk/acrc}. We would like to thank Dr Conor Fitzpatrick (University of Manchester) and Dr Mika Vesterinen (University of Warwick) for their careful reading of an earlier draft of this manuscript, and their helpful comments. This research was supported by the Science and Technology Facilities Research Council, UK.

\bibliographystyle{JHEP}
\bibliography{bibliography}

\end{document}